\RequirePackage{snapshot}
\documentclass[letterpaper,twocolumn,10pt,times]{article}

\usepackage{wrapfig}
\usepackage{comment}
\usepackage{colortbl}

\usepackage{hyperref}
\usepackage{tabularx}
\def\UrlBreaks{\do\/\do-}

\usepackage{color,soul}
\usepackage{graphics}
\usepackage{hyperref}
\usepackage{lipsum}

\usepackage{tikz}
\usetikzlibrary{decorations.pathmorphing, patterns,shapes,snakes} % for zigzag, etc.

\usepackage[inline]{enumitem}	% for fine control of list env. support inline enumerate. 
\usepackage{listings} % for code listing
\usepackage{booktabs}	% xzl: for fancy tables
\usepackage{lipsum}

\usepackage{capt-of}	% xzl: used to override the ``figure'' and ``table'' in caption.

\usepackage{ulem}  % xzl: for strikeout, which seems more robust than \st{} provided by soul package. 

\ifdefined\HCode
\usepackage[compatibility=false]{caption}

\else
\usepackage{subcaption} 
\fi

\RequirePackage{titling}

\usepackage{titlesec}
\pretitle{\begin{center}
\Large\bfseries}
\preauthor{\begin{center}\normalfont\large%
    \begin{tabular}[t]{c}}
\postauthor{\end{tabular}\par\end{center}}
\predate{}
\postdate{}

\usepackage{mathrsfs}	% xzl: for Ralph Smith’s Formal Script Font (rsfs)
\usepackage{amsfonts}
\usepackage{inconsolata} % xzl: this uses inconsolata for texttt

\usepackage{mathtools}% Loads amsmath. for underbrace

\usepackage[export]{adjustbox} % xzl: flush figure to left/right

\definecolor{airforceblue}{rgb}{0.36, 0.54, 0.66}
\newcommand*\circled[1]{\tikz[baseline=(char.base)]{
		\node[shape=circle,fill=airforceblue,draw,text=white,inner sep=1pt] (char) {#1};}}

\newcommand*\snakezigzag		
{\tikz
\draw[snake=zigzag,thick] (0,1.5) -- (1,1.5);
}

\newcommand*\snakebump
{\tikz
\draw[snake=bumps,thick] (0,1.5) -- (1,1.5);
}

\newcommand*\invsnakebump
{\tikz
\draw[snake=bumps,thick,rotate=180] (0,1.5) -- (1,1.5);
}
		
\newcommand{\Paragraph}[1]{\vskip 3pt\noindent\textbf{#1 }}	 % used to be 6pt

\newcommand{\textbfit}[1]{\textbf{\textit{#1}}}

\renewcommand{\paragraph}[1]{\vskip 3pt\noindent\textbf{#1 }}	 % used to be 6pt

\renewcommand{\st}[1]{\sout{#1 }}	 % st{} from soul not robust enough

  {\begin{list}{$\bullet$}%
     {\setlength{\parsep}{0pt}%
      \setlength{\topsep}{0pt}%
      \setlength{\itemsep}{2pt}}}%
  {\end{list}}
\newcommand\note[1]{\sethlcolor{yellow} \hl{#1}} % highlighted notes of other colors.
\newcommand\noted[1]{} % remove highlights.

\newcommand\todo[1]{\sethlcolor{ForestGreen} \hl{todo: #1}} % highlighted notes of other colors.

\newcommand\sz[1]{{\color{violet}{sz: #1}}}
\newcommand\lwg[1]{{\color{Emerald}
		{lwg: #1}}} % highlighted notes of other colors.

\newcommand\sect[1]{Section~\ref{sec:#1}}	% NB: does not play well with \note{}
		
\newenvironment{myitemize}%
  {\begin{itemize}
	[leftmargin=0cm,
		itemindent=.3cm,
		labelwidth=\itemindent,
		labelsep=0pt,
		parsep=3pt,
		topsep=2pt,
		itemsep=1pt,
		align=left]
  }%
  {\end{itemize}}    

\newenvironment{myenumerate}%
  {\begin{enumerate}
	[leftmargin=0cm,itemindent=.5cm,labelwidth=\itemindent,
		labelsep=0pt,
		parsep=1pt,
		topsep=1pt,
		itemsep=3pt,
		align=left]
  }%
  {\end{enumerate}}    
  
\newenvironment{myenumerateinline}%
  {\begin{enumerate*}
	[label=\roman*)]
  }%
  {\end{enumerate*}}      

\newcommand{\code}[1]{\texttt{\small{#1}}}	
\newcommand{\sys}{ARK}

\newcommand{\pmcore}{peripheral core}
\newcommand{\Pmcore}{Peripheral core}

\newcommand{\lwk}{peripheral kernel}

\newcommand{\snr}{suspend/resume}

\newcommand{\iophase}{device phase}

\newcommand{\ekc}{emulated services}

\newcommand{\sysmodel}{transkernel}
\newcommand{\Sysmodel}{Transkernel}

\newcommand{\caveat}{\noindent \textit{Caveats fixed} \hspace{1mm}}

\makeatletter
\def\@copyrightspace{\relax}
\makeatother

\usepackage{usenix2019_v3}
\usepackage{epsfig,endnotes}	
\usepackage{url}
\usepackage{multirow}
\usepackage{authblk}
\usepackage{threeparttable}
\begin{document}

%don't want date printed
\date{}

\title{\vspace{10pt}\Large \bf Transkernel: Bridging Monolithic Kernels to Peripheral Cores}
\title{\Large \bf Transkernel: Bridging Monolithic Kernels to Peripheral Cores}

%\author{%
%Author A\affmark[1], Author B\affmark[1], Author C\affmark[1], Author %D\affmark[2], and Author E\affmark[2]\\
%\affaddr{\affmark[1]Department of Computer Science}\\
%\affaddr{\affmark[2]Department of Mechanical Engineering}\\
%\email{\{A,B,C,D,E\}@university.edu}\\
%\affaddr{\LaTeX\ University}%
%}

%\author{
%\rm Hongyu Miao\affmark[1], \rm Heejin Park\affmark[1], \rm Myeongjae Jeon\affmark[2], \\
% \rm Gennady Pekhimenko\affmark[2], \rm Kathryn S. McKinley\affmark[3], and \rm Felix Xiaozhu Lin\affmark[1]\\
%\affaddr{\affmark[1]Purdue ECE~~~~~~~~~\affmark[2]Microsoft Research ~~~~~~~~~~\affmark[3]Google}
%}

%for single author (just remove % characters)
%\author{
%{\rm Your N.\ Here\affmark[1]}\\
%\affmark[1]Your Institution
%\and
%{\rm Second Name\affmark[2]}\\
%\affmark[2]Second Institution
% copy the following lines to add more authors
% \and
% {\rm Name}\\
%Name Institution
%} % end author

% For camera-ready
%\author{
%	{\rm Liwei Guo}
	%\thanks{Both authors contribute equally to this %work}\\
%	\affaddr{Purdue ECE}
%	\and
%	{\rm Shuang Zhai \textsuperscript{*}}\\
%	\affaddr{Purdue ECE}
%	\and
%	{\rm Yi Qiao}\\
%	\affaddr{Purdue ECE}
%	\and
%	{\rm Felix Xiaozhu Lin}\\
%	\affaddr{Purdue ECE}	
%}

\author[]{Liwei Guo}
\author[]{Shuang Zhai}
\author[]{Yi Qiao}
\author[]{Felix Xiaozhu Lin}
\affil[]{Purdue ECE}

\maketitle

%\vspace*{-10em}
% Use the following at camera-ready time to suppress page numbers.
% Comment it out when you first submit the paper for review.
%\pagenumbering{gobble}
% !TeX root = main.tex

\subsection*{Abstract}

%Short-lived tasks have a large impact on the battery life of mobile/embedded platforms. 
%%In executing such a task, the OS kernel drives the whole platform out and in deep sleep. 
%%It spends a dominating portion of energy on device suspend/resume, an execution phase in which the kernel operates various IO devices and waits for their power state transitions. 
%%In executing such tasks, the platform spends a dominant portion of energy 
%%%an execution phase in which the kernel operates various IO devices and orchestrates their power state transitions. 
%%%an OS kernel execution phase operating various IO devices and orchestrating their power state transitions. 
%%when the OS kernel operates various IO devices and orchestrates their power state transitions. 
%For such tasks, a dominant portion of energy is consumed by the kernel execution phase
%%an execution phase in which the kernel operates various IO devices and orchestrates their power state transitions. 
%%an OS kernel execution phase operating various IO devices and orchestrating their power state transitions. 
%for operating various IO devices and orchestrating their power state transitions. 

Smart devices see a large number of ephemeral tasks driven by background activities.
%Before and after executing such a task, the OS kernel wakes up the platform and puts it to sleep, 
In order to execute such a task, the OS kernel wakes up the platform beforehand and puts it back to sleep afterwards. 
In doing so, the kernel operates various IO devices and orchestrates their power state transitions. 
%Such kernel execution phases have high energy impact and are difficult to optimize.
Such kernel executions are inefficient as they mismatch typical CPU hardware.
They are better off running on a low-power, microcontroller-like core, i.e., \pmcore{}, relieving CPU from the inefficiency.
%We advocate for executing these kernel phases with a low-power, microcontroller-like core, dubbed \pmcore{},  on behalf of the CPU, hence leaving the CPU off.

%To minimize the energy impact, we advocate offloading the kernel's device suspend/resume from CPU to a low-power, microcontroller-like core, dubbed \pmcore{}, hence leaving the CPU off. 
%To minimize the energy impact, we advocate using a low-power, microcontroller-like core, dubbed \pmcore{}, to execute such a kernel phase on behalf of the CPU, hence leaving the CPU off. 

%To this end, 
%To enable a \pmcore{} to execute a commodity kernel,
%Yet, to execute the code from a complex, fast-evolving commodity kernel (e.g. Linux) on a \pmcore{},
%Yet, for a \pmcore{} to execute a monolithic kernel (e.g., Linux), existing approaches either incur high engineering effort or high overhead. 

%We take a radical approach with a new executor model called \textit{\sysmodel{}}.
We therefore present a new OS structure, in which a lightweight virtual executor called \textit{\sysmodel{}} offloads specific phases from a monolithic kernel. 
The \sysmodel{} translates stateful kernel execution through cross-ISA, dynamic binary translation (DBT);
it emulates a small set of stateless kernel services behind a narrow, stable binary interface; 
it specializes for hot paths; 
it exploits ISA similarities for lowering DBT cost. 

Through an ARM-based prototype, we demonstrate \sysmodel{}'s feasibility and benefit.
%Overall, we contribute a new OS structure that incorporates cross-ISA DBT and ABI emulation for harnessing a heterogeneous SoC.
We show that while cross-ISA DBT is typically used under the assumption of efficiency \textit{loss}, it can enable efficiency \textit{gain}, 
even on off-the-shelf hardware. 
%%even atop off-the-shelf heterogeneous SoCs. 
%even atop off-the-shelf hardware. 

%it demonstrates that while cross-ISA DBT is typically used in trading efficiency for software portability, it can gain efficiency through careful software optimizations, even atop off-the-shelf hardware. 
%it demonstrates that while cross-ISA DBT is typically used in sacrificing efficiency for software portability, it can be used to gain efficiency, 
%our result demonstrates that while cross-ISA DBT is typically used for trading efficiency for software portability, the technique can gain efficiency, 
%Our result demonstrates that while cross-ISA DBT is typically used at the cost of efficiency, it can be used to \textit{gain} efficiency, 
%%even atop off-the-shelf heterogeneous SoCs. 
%even atop off-the-shelf hardware. 
%
% -- good --
%Our result demonstrates that while cross-ISA DBT is typically used under the assumption of efficiency \textit{loss}, it can be a key enabler for efficiency \textit{gain}, 
%even atop off-the-shelf hardware. 

%\input{todo}
% !TeX root = main.tex

\section{Introduction}
\label{sec:intro}

%Modern mobile/wearable/IoT devices often run a large number of ephemeral tasks. 
Driven by periodic or background activities,
modern embedded platforms\footnote{This paper focuses on battery-powered computers such as smart wearables and smart things. 
They run commodity OSes such as Linux and Windows. We refer to them as embedded platforms for brevity.} often run a large number of ephemeral tasks. 
%These tasks are intermittent and short-lived, driven by background software activities. 
Example tasks include acquiring sensor readings, refreshing smart watch display~\cite{understanding-android-wear}, push notifications~\cite{drowsy-pm}, and periodic  data sync~\cite{xu2013mobisys}.
%Such an ephemeral task typically lasts from hundreds of ms~\cite{understanding-android-wear,drowsy-pm} to a few seconds~\cite{xu2013mobisys}; 
They drain a substantial fraction of battery, e.g., 30\%  for smartphones~\cite{backgnd-in-wild,backgnd-in-wild2} and smart watches~\cite{mobisys17-wear}, 
and almost the entire battery of smart things for surveillance~\cite{farmbeats}.
% https://www.sciencedirect.com/science/article/pii/S1389128616000086#fig0005
%e.g. 29\% on smartphones~\cite{backgnd-in-wild} and 27\% of smart watches~\cite{mobisys17-wear}. 
%To execute such a task, the whole system often exits from a deep-sleep state to run user tasks briefly, and re-enters the deep-sleep state. 
%To execute such an ephemeral task, the whole platform exits from deep sleep, executes user code, and re-enters deep sleep afterwards. 
To execute an ephemeral task, a commodity OS kernel, 
typically implemented in a monolithic fashion,
 drives the whole hardware platform out of deep sleep beforehand (i.e., ``resume'') and puts it back to deep sleep afterwards (i.e., ``suspend''). 
%Being a core kernel functionality for almost two decades, the 
During this process, the kernel consumes much more energy than the user code~\cite{understanding-android-wear}, up to 10$\times$ shown in recent work~\cite{drowsy-pm}.
%kernel procedure can consume much more energy than the ephemeral task itself~\cite{liu16mobisys}, up to 10$\times$~\cite{drowsy-pm}.

%This \textit{\snr{}} procedure~\cite{suspend-to-ram} has been an integral part of OS kernel power management for almost two decades, and now unfortunately becomes a heavy burden for ephemeral tasks. 
%In such a procedure, the execution of a commodity OS kernel becomes a heavy burden.
%Unfortunately, 
%It has been show that ephemeral tasks are burdened with the procedure of waking up the platform and putting it back to sleep:
%the execution of a commodity OS kernel:

%Such a procedure is burdened with the execution of a commodity OS kernel.
% https://elinux.org/Device_Power_Management_Specification
%\SnR{} is known to be slow, each taking hundreds ms to seconds~\cite{XXX,drowsy-pm}.
%Yet, they contribute a significant, sometimes dominating, portion, to the total system energy, in some scenarios, the portion is as high as 90\%~\cite{drowsy-pm}.
% --- drowsy, figure 9 --- 

\begin{figure}[t]
    \centering
   	\includegraphics[width=0.48\textwidth]{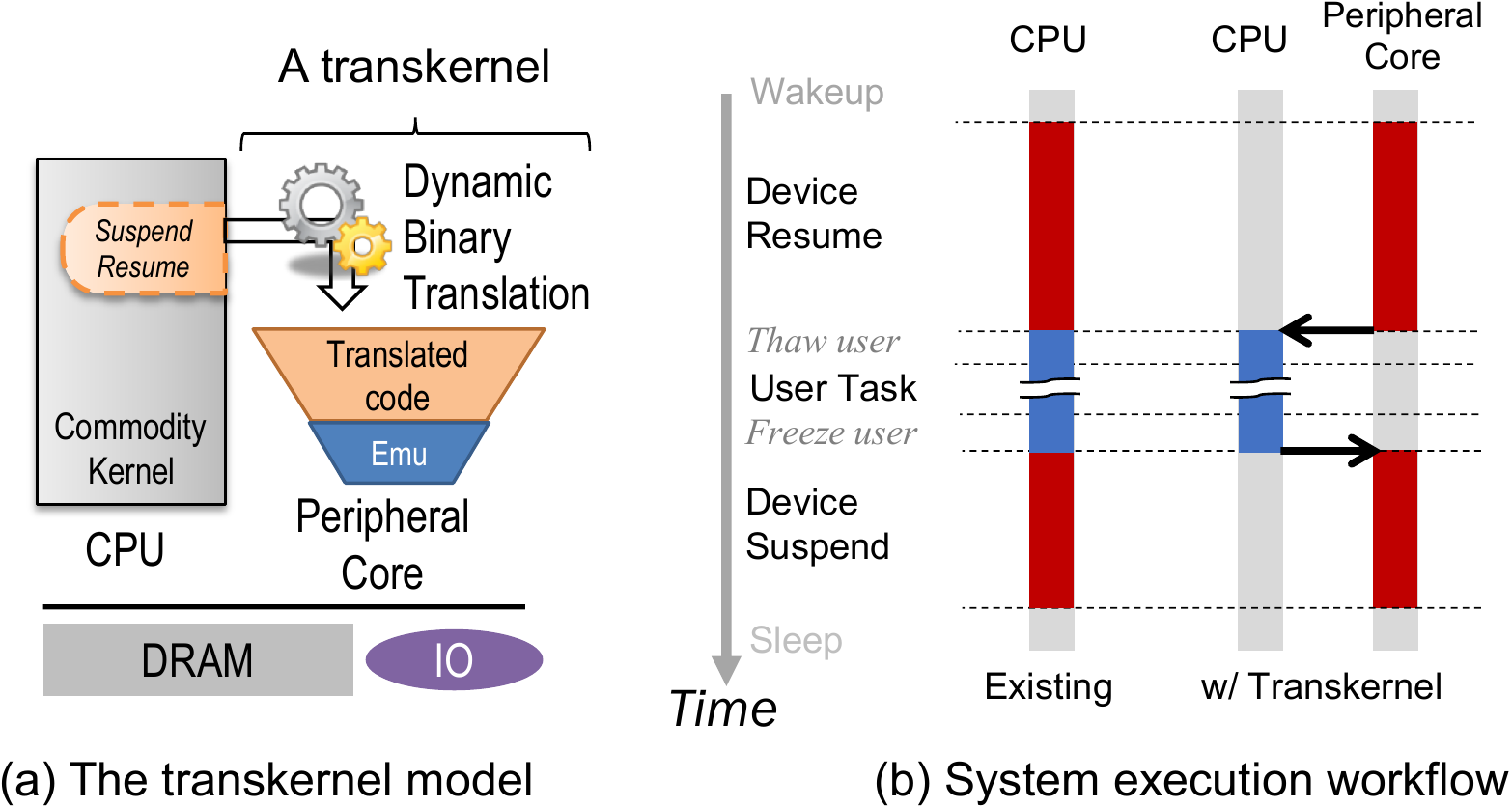}
    \caption{An overview of this work}
    \label{fig:offload}
    \vspace*{-15pt}
\end{figure}

%Why is \snr{} so expensive? 
Why is the kernel so inefficient? 
Recent studies~\cite{linaro-sdio-pm,zhai17hotmobile,understanding-android-wear}
%as confirmed with new evidences in Section~\ref{sec:bkgnd}, 
show the bottlenecks as two kernel phases called \textit{device \snr{}} as illustrated in Figure~\ref{fig:offload}. 
In the phases, the kernel operates a variety of IO devices (or \textit{devices}  for brevity). 
It invokes device drivers, cleans up pending IO tasks, and ensures devices to reach expected power states. 
%the kernel execution is choked by IO devices (we refer to them as \textit{devices} in this paper).
%spends substantial (?) time waiting for IO power state transitions: 
%To suspend/resume an embedded platform, the kernel typically operates tens of different devices by invoking their respective drivers, finishes pending IO tasks, and ensures these devices to reach the expected power states. 
The phases encompass concurrent execution of drivers, deferred functions, and hardware interrupts; 
they entail numerous CPU idle epochs; 
their optimization is proven difficult (\S\ref{sec:bkgnd}) ~\cite{zhai17hotmobile,linux-asyncpm,lkml-pm}. 
%Such IO waits are proven difficult to reduce (\S\ref{sec:bkgnd}) ~\cite{zhai17hotmobile,linux-asyncpm,lkml-pm}. 

%Extensive prior work suggests 
%Due to these characteristics, 
%As an IO-bound execution kernel execution,
%As an IO-bound phase in the kernel execution, \note{improve}
%As an IO-intensive execution phase, \note{improve}

% !TeX root = main.tex

%\begin{figure}[t!]
%	
%	\begin{minipage}[b]{0.22\textwidth}
%		\includegraphics[height=1.1\textwidth]{figs/arch-ours}
%		        \subcaption{The \sysmodel{} model}
%%		\label{fig:eval:read}
%	\end{minipage}
%	\qquad
%%	\begin{minipage}[b]{0.22\textwidth}
%%		\includegraphics[angle=90,height=1.1\textwidth]{figs/workflow-scratch}
%%		        \subcaption{Execution timeline}
%%		\label{fig:eval:write}
%%	\end{minipage}	
%	\begin{minipage}[b]{0.12\textwidth}
%		\includegraphics[height=1.1\textwidth]{figs/workflow}
%		        \subcaption{Execution timeline. \textbf{remove details (syncfs) highlight dev sr}}
%%		\label{fig:eval:write}
%	\end{minipage}	
%	
%	\caption{An overview of this work \textbf{first resume, then sleep?}}
%%	\vspace{-15pt}		% use as needed
%	\label{fig:overview}
%\end{figure}

We deem that device \snr{} mismatches CPU.  
It instead would be much more efficient on low-power, microcontroller-like cores, as exemplified by ARM Cortex-M.
%To relieve CPU from the burden, we deem that device \snr{} should be executed by low-power, microcontroller-like cores. 
%such as simple pipelines and small caches.
These cores are already incorporated as \textit{\pmcore{}}s on a wide range of modern system-on-chips (SoCs) used in production such as Apple Watch~\cite{apple-watch-teardown} and Microsoft Azure Sphere~\cite{mediatek-mt3620}. 
%As exemplified by ARM Cortex-M, such cores have trimmed-down ISAs and simple microarchitectures.
%For workloads with heavy IO and low performance demands, 
On IO-intensive workloads, a \pmcore{} delivers is much more efficient than the CPU due to lower idle power and higher execution efficiency~\cite{reflex,k2,dreamweaver,somniloquy}. 
%, it can execute device \snr{} operates autonomously while keeping CPU off;
%Note that running \textit{user programs} on \pmcore{}s is a non-goal: doing so often yields marginal benefits due to sparse idle epochs; it would require support POSIX functions requires (full-fledged) kernel, which is difficult and out of scope.
%As shown in Figure~\ref{fig:offload}(b), 
Note that running \textit{user code} 
(which often builds atop POSIX)
on \pmcore{}s is a non-goal:
on one hand, doing so would gain much less efficiency due to fewer idle epochs in user execution;
on the other hand, doing so requires to support a much more complex POSIX environment on \pmcore{}s. 
%require to support a POSIX environment, a much more complex stack than running specific kernel phases. 

%because as these programs often build on POSIX or Android.
%We consider such an effort orthogonal to this work.
%\note{here since translation is not introduced yet, I choose to say ``offload'' POSIX functions. However I'm not quite sure if it is a good idea to call out difficulties of implementation considering it is already done by Jon Howell's POSIX emulation.}

%on the same SoC~\cite{reflex,k2,dreamweaver,somniloquy}. 
%A \pmcore{} is often loosely coupled with the main CPU; 

%A \pmcore{} often resides in a separate power domain; 
%it is capable of operate autonomously while keeping the CPU off. 

%the resultant interfaces between the kernel and the 

%Offloading the \textit{kernel execution phase} of device \snr{} from the CPU to a \pmcore{} is challenging. 
Offloading the execution of a commodity, monolithic kernel raises practical challenges, not only
i) that the \pmcore{} has a different ISA and wimpy hardware 
but also 
ii) that the kernel is complex and rapidly evolving~\cite{cocinelle1}.
Many OS proposals address the former while being inadequate in addressing the latter~\cite{popcorn,helios,multikernel,rainforest,legoos}.
%Notably, a multikernel could span its execution over CPU and a \pmcore{};
%it, however, depends on a wide \textit{binary interface} (ABI) for synchronizing the kernel states between the two ISAs. 
For instance, one may refactor a monolithic kernel to span it over CPU and a \pmcore{};
the resultant kernel, however, depends on a wide binary interface (ABI) for synchronizing state between the two ISAs. 
This interface is brittle. 
%and depends on both kernel configuration and source code. 
%as the kernel changes build configurations and its source rapidly evolves~\cite{cocinelle1}.
%\note{say that have to maintain one software for each commodity kernel build}
As the upstream kernel evolves,
maintaining \textit{binary compatibility} across different ISAs inside the kernel itself soon becomes unsustainable.
%, an effort barely sustainable. 
Instead, we argue for the code running on \pmcore{}s to enjoy 
firmware-level compatibility:
%good compatibility just as firmware:
developed and compiled \textit{once}, it should work with \textit{many} builds of the monolithic kernel -- generated from different configurations and source versions.

%--- verbose --- 
\begin{comment}
%We seek to make the offloading practical for a complex, fast-evolving kernel. 
We seek to enable the offloading with the following goals: 
%i) making engineering effort tractable and especially reusing much of the kernel code;
i) tractable engineering effort with good reuse of the commodity kernel code;
ii) developing and compiling software for a \pmcore{} \textit{once}; 
running the software with \textit{many} builds of the kernel -- generated from different configurations and source versions;
%ii) the software for the \pmcore{} is developed and built \textit{once} and it can run with \textit{many} builds of the kernel -- generated from different configurations and source releases;
iii) low runtime overhead. 
\end{comment}

Our response is a radical design called \textit{\sysmodel{}}, 
a lightweight virtual executor empowering a \pmcore{} to run specific kernel phases -- device \snr{}. 
Figure~\ref{fig:offload} overviews the system architecture.
A \sysmodel{} executes unmodified kernel binary through cross-ISA, dynamic binary translation
(DBT), a technique previously regarded as expensive~\cite{popcorn} and never tested on microcontroller-like cores to our knowledge.
%runs a DBT engine, which translates the unmodified kernel binary for most code executed in the offloaded kernel phases; 
Underneath the translated code,
a small set of emulated services act
%with lightweight internals and as drop-in replacement for their counterparts in the kernel.
as lightweight, drop-in replacements for their counterparts in the monolithic kernel.
Four principles make \sysmodel{} practical:
%To make the whole design feasible, we follow four principles:
\begin{myenumerateinline}
\item translating stateful code while emulating stateless kernel services; 
%\item minimizing dependency on the monolithic kernel;
%ii) minimizing coupling with the commodity kernel;
%ii) minimizing coupling between cores;
%ii) choosing narrow and stable interfaces for emulation; 
\item identifying a narrow, stable translation/emulation interface; 
\item specializing for hot paths; 
\item exploiting ISA similarities for DBT.
\end{myenumerateinline}

%We demonstrate the \sysmodel{} model by building a prototype called \sys{}.
We demonstrate a \sysmodel{} prototype called \sys{} (\textbf{A}n a\textbf{R}m trans\textbf{K}ernel).
Atop an ARM SoC, \sys{} runs on a Cortex-M3 peripheral core (with only 200 MHz clock and 32KB cache) alongside Linux running on a Cortex-A9 CPU.
%the SoC incorporates Cortex-A9 cores as the CPU which runs Linux.
%\sys{} meets our goals.
\sys{} transparently translates unmodified Linux kernel drivers and libraries.
It depends on a binary interface consisting of only 12 Linux kernel functions and one kernel variable, which are stable for years.
\sys{} offers complete support for device \snr{} in Linux, 
capable of executing diverse drivers that implement rich functionalities (e.g., DMA and firmware loading) and invoke sophisticated kernel services (e.g., scheduling and IRQ handling).
As compared to native kernel execution, 
\sys{} only incurs 2.7$\times$ overhead, 5.2$\times$ lower than a baseline of off-the-shelf DBT.
%commodity cross-ISA DBT~\cite{qemu}.
%The low cost makes \sys{} beneficial:
\sys{} reduces system energy by 34\%, resulting in tangible battery life extension under real-world usage.
%we estimate to extend battery life in real-world usage by XX\%. 

We make the following contributions on OS and DBT:
\begin{myitemize}
%\item We present the \sysmodel{} model and its principles for a \pmcore{} to execute the kernel \snr{} path. 
%\item We present the \sysmodel{} model for executing a phase in a commodity kernel on a \pmcore{}. 
\item We present the \sysmodel{} model.
%in which a \pmcore{} executes specific phases of a monolithic kernel on behalf of CPU.
%To the design space of OS for heterogeneous multi-processors, the \sysmodel{} contributes a new design point that combines DBT for transparent code reuse and state sharing and a miniature runtime (?).
In the design space of OSes for heterogeneous multi-processors, the \sysmodel{} represents a novel point: it combines DBT and emulation 
for bridging ISA gaps and for catering to core asymmetry, respectively.
%for tractable engineering.

\item We present a \sysmodel{} implementation, \sys{}. 
Targeting Linux, \sys{} presents specific tradeoffs between kernel translation versus emulation;
it identifies a narrow interface between the two;
it contributes concrete realization for them.

% Liwei's
%\sys{} brings kernel tasks to the peripheral core: compared with refactoring existing code which has to reason about hundreds of data types and interfaces, \sys{} reduces them to a few, identifies specific tradeoffs between kernel translation versus emulation, and contributes concrete interfaces and realization for them.

% --- verbose ---- 
%\sys{} meets our goal of tractable engineering efforts and ``build once run with many''. 
%\sys{} offers tangible benefits of energy efficiency. 

% --- useful? ---
% To our knowledge, it features the first cross-ISA DBT engine running atop a microcontroller-like core. 
\item
Crucial to the practicality of \sys{}, 
we present an inverse paradigm of cross-ISA DBT, in which a microcontroller-like core translates binary built for a full-fledged CPU. 
% executes unmodified binary code on behalf of a resource-rich CPU. 
We contribute optimizations that systematically exploit ISA similarities.
%Beyond \snr{}, 
Our result demonstrates that while cross-ISA DBT is typically used under the assumption of efficiency \textit{loss}, it can enable efficiency \textit{gain}, 
even on off-the-shelf hardware. 

%that while cross-ISA DBT is typically for trading efficiency for other objectives\note{cite}, it can \textit{gain} efficiency through careful software optimization, even atop off-the-shelf hardware.
\end{myitemize}

%\note{modify this}
The source code of \sys{} can be found at \url{http://xsel.rocks/p/transkernel}.
\section{Motivations}
\label{sec:bkgnd}

%We next describe OS \snr{}, the kernel execution that dominates ephemeral tasks \improve{}.  
%We discuss its low efficiency, for which we advocate to mitigate by offloading to the \snr{} path a \pmcore{}. 
%To this end, we examine existing software structures and their inadequacies. 

%We next present observations on \snr{}. 
%We show its primary bottleneck, the \iophase{}, can be significantly mitigated by offloading to a microcontroller-profile processor on the same SoC. 

We next discuss device \snr{}, the major kernel bottleneck in ephemeral tasks, and that it can be mitigated by running on a \pmcore{}.
%We argue to mitigate its inefficiency with a \pmcore{}. 
We show difficulties in known approaches and accordingly motivate our design objectives. 

%By profiling the Linux kernel on XXX platforms, we add results to the observation that slow \iops{}s are the major causes of \snr{} inefficiency. 
%We analyze the \snr{} code, showing its complexity and rapid evolution. 
%We advocate offloading OS \snr{} to an extremely simple, low-power core.
%%These results motivate to offload OS \snr{} to an extremely simple, efficient core.

%\subsection{Observations on OS suspend/resume}
%\subsection{Observations on device suspend/resume}
\subsection{Kernel in device suspend/resume}
\label{sec:bkgnd:code}

Expecting a long period of system inactivity, an OS kernel 
puts the whole platform into deep sleep:
%shut down CPU cores and most IO devices and transits the whole platform (machine) into a deep sleep state.
%puts its platform into a deep sleep state 
%This suspend process is often initiated by userspace. 
in brief, the kernel synchronizes file systems with storage, 
freezes all user tasks, 
turns off IO devices
%\footnote{We refer to them as \textit{devices} for simplicity} 
(i.e., device suspend), 
and finally powers off the CPU. 
%On a successful execution suspend, function calls will never return. \note{improve}
To wake up from deep sleep, the kernel performs a mirrored procedure~\cite{suspend-to-ram}. 
%\note{For a detailed description, see Linux documentation}~\cite{suspend-to-ram}.
%This is reflected in the execution time of in an ephemeral task: 
In a typical ephemeral task, the above kernel execution takes hundreds of milliseconds~\cite{intel-suspendresume} while the user execution often takes tens of milliseconds~\cite{understanding-android-wear}; 
the kernel execution often consumes several times more energy than the user execution~\cite{drowsy-pm}.
\paragraph{Problem: device \snr{}}
By profiling recent Linux on multiple embedded platforms, 
our pilot study~\cite{zhai17hotmobile} shows the aforementioned kernel
% execution is bottlenecked at the phases of device \snr{}, in which 
%the kernel finishes pending IO tasks and drives transitions of device power states. 
 execution is bottlenecked by device \snr{}, in which
% the kernel phases right before powering off CPU and right after powering on CPU.  
%the kernel phases right before powering off the CPU and right after powering on the CPU.
the kernel cleans up pending IO tasks and manipulates device power states. 
The findings are as follows.
%We briefly summarize the prior findings below. 
%more specifically, the slow power state transitions of IO devices. 
%They frequently keeps CPU waiting. 
%We next briefly summarize the prior findings with new evidences in Figure~\ref{fig:suspend}.
%The prior work reports the following findings, as confirmed with new evidences in Figure~\ref{fig:suspend}. 
%
\begin{myenumerateinline}
\item 
%The kernel executes device drivers and the needed services, finishing pending IO tasks and driving power state transitions of these devices.
\textbf{\textit{Device \snr{} is inefficient.}} 
It contributes 54\% on average and up to 66\% to the total kernel energy consumption.
%The transitions of device power states take long, contributing 68\% of time in device \snr{}. 
%The transitions of device power states take long.
CPU idles frequently in numerous short epochs, typically in milliseconds.
%CPU waits frequently.
%The waits occur in numerous short epochs (typically in milliseconds), rather than a few long epochs. 
% idling or spinning. 
% --- useful. --- 
%The total device waits contribute to 14\% of the energy spent on \snr{}.
\item 
%The IO waits due to power state transitions are contributed 
%During device \snr{} on an embedded platform, the kernel often operates tens of different devices.
\textbf{\textit{Devices are diverse.}}
On a platform, the kernel often suspends and resumes tens of different devices.
%Across platforms, different devices incur long kernel executions. 
Across platforms, the bottleneck devices are different. 
%Across platforms, the devices incurring long kernel executions are different. 
%the top energy-hungry devices are different. 
%The waits are contributed by diverse IO devices on the profiled embedded platforms.
%The waits are contributed by a variety of IO devices on the profiled embedded platforms.
% --- useful --- 
%These discourage point solutions to specific device drivers or specific wait epochs.  
\item 
%Optimizing device \snr{} is difficult. 
\textbf{\textit{Optimization is difficult.}} 
%i) \iops{}s are bound by slow peripherals and low-speed buses commonly seen on embedded platforms, as well as physical limits (e.g. voltage ramping up). 
%i)
Device power state transitions are bound by slow hardware and low-speed buses, as well as physical factors (e.g., voltage ramp-up). 
% --- the following is already covered --- %
%some wait regions cannot be parallelized, e.g. claiming emmc host which waits for any outstanding requests to complete. (Can't this be overlapped with waits for other IO?)
%Second, it is difficult to overlap the power state transitions of multiple IO devices. 
%ii) 
%Overlapping the state transitions of different devices sees limited efficacy: 
%Traditionally, IO devices are suspended synchrounously, in the order of when they are brought up during system boot time.
%This may alleviate the problem but cannot eliminate it. 
While Linux already parallelizes power transitions with great efforts~\cite{linux-asyncpm, lkml-pm},
many power transitions must happen sequentially per \textit{implicit} dependencies of power, voltage, and clock. 
%Consider a simplified example: if i) the kernel simultaneously turns off a voltage regulator and an SD card and ii) the regulator supplies power to the SD card, the SD card may lose dirty data that are yet to be written back.
%In practice, inter-device dependency is more subtle and often implicit, for which the OS has no full knowledge.
%Due to such a widespread concern and after a long debate~\cite{linux-asyncpm}, 
%Modern Linux already overlaps limited asynchrony in modern Linux kernels~\cite{linux-asyncpm, lkml-pm}, 
%Modern Linux already overlaps device power state transitions~\cite{linux-asyncpm, lkml-pm}. 
%Yet, as shown in Figure~\ref{fig:suspend}, 
%As a result, CPU idles for up to 68\% of the duration of device \snr{}.
As a result, CPU idle constitutes up to 68\% of the device \snr{} duration.
%\st{Prior research explored lazy suspend/resume of devices~\cite{drowsy-pm}; it however never entered the mainline Linux, as will be discussed in \sect{related}}.
% due to dependency difficulty to our knowledge.
 
%The raised ideas, e.g. lazily resuming, have been explored [34] but never enter the mainline Linux to our knowledge. While acknowledging the value of such kernel optimizations, we believe ARK is a key complement that works on unmodified binaries. ARK can co-exist with the mentioned optimizations in the same kernel. 

%major wait epochs still cannot be overlapped. 

%However, it is less effective on SoC that has a great variety of devices (and thus drivers). 
%Many drivers opts out for being suspended/resumed in parallel to other devices. 
%As a result, such ``intra-driver'' asynchrony is more effective on workstations or servers, which often embrace multiple SATA disks or USB devices. 
%However, it is less effective on SoC that has a great variety of devices (and thus drivers). 

%Note that even if the kernel perfectly parallelizes all IO PM (a highly unlikely situation), the whole suspend/resume is still bottlenecked by multiple slow IO devices, each taking tens to hundreds of ms (XXX). 
%This is still far from the satisfactory (10 ms for servers~\cite{powernap} and sub ms for wearables).
\end{myenumerateinline}

\paragraph{Challenge: Widespread, complex kernel code}
%Since IO is the primary inefficiency, 
%Since IO is the dominant inefficiency in \snr{}, we inspect the IO-related code. 
Device \snr{} invokes multiple kernel layers~\cite{cocinelle1,understanding-driver}.
Specifically, it invokes functions in individual drivers (e.g., MMC controllers), driver libraries (e.g., the generic clock framework), kernel libraries (e.g., for radix trees), and kernel services (e.g., scheduler).
In a recent Linux source tree (4.4), 
we find that over 1000 device drivers, which represent almost all driver classes, 
implement \snr{} callbacks in 154K SLoC.
These callbacks in turn invoke over 43K SLoC in driver libraries, 8K SLoC in kernel libraries, and 43K SLoC in kernel services. 
The execution is control-heavy, with dense branches and callbacks.

\paragraph{Opportunities}
We observe the following kernel behaviors in device \snr{}. 
%Fortunately, the following kernel behaviors create opportunities for us to address the inefficiency. 
%aforementioned 
\begin{myenumerateinline}

\item %\paragraph{Insensitivity to execution time}
\textbf{\textit{Low sensitivity to execution delay}} \hspace{1mm}
On embedded platforms, most ephemeral tasks are driven by background activities~\cite{drowsy-pm,powernap,backgnd-in-wild}.
This contrasts to many servers for interactive user requests~\cite{sw-engagement,powernap}. 
%\note{highlight this up front}
	
\item 
\textbf{\textit{Hot kernel paths}} \hspace{1mm}
%A beaten path is the kernel code exercised during a successful \snr{} cycle. 
%Most \snr{} execution the kernel exercises
In successful \snr{}, the kernel acquires all needed resources and encounters no failures~\cite{lock-in-pop}. 
Off the hot paths, the kernel handles rare events such as races between IO events, resource shortage, and hardware failures.
These branches typically cancel the current \snr{} attempt, perform diagnostics, and retry later. 
Unlike hot paths, they invoke very different kernel services, e.g., syslog.

%For this reason, unbeaten paths often invoke much more diverse libraries and kernel infrastructure functions (e.g. file system or syslog) than the beaten paths. 
%\todo{show numbers; more examples; mention that beaten paths execute a small set of kernel functions?}

\item 
%\paragraph{A simple concurrency model}
\textbf{\textit{Simple concurrency}} 
exists among the syscall path (which initiates \snr{}), interrupt handlers, and deferred kernel work. 
%In particular, deferred work, as exemplified by Linux's work items and tasklets, are extensively used to overlap IO latencies. 
The concurrency is for hardware asynchrony and kernel modularity rather than exploiting multicore parallelism.
%all user tasks are frozen and the remaining execution is bound by IO. 

% the concurrency is for asynchrony rather than exploiting multicore parallelism. 

%, e.g. to schedule a future check of a device state, for which it invokes XXXX (tasklet functions). 
%For instance, the \snr{} path iterates through drivers, requesting all device-specific callbacks to be executed asynchronously, and executes them later. 

%This contrasts to user-facing servers that are more sensitive to wake up latency~\cite{sw-engagement,powernap}. 

%Certain external events may demand fast resume, e.g. a user unlocking the smartphone screen, for which we will consider XXX. 
%This allows to achieve higher efficiency in \snr{} at the price of increased execution time. 
%This allows the OS to trade low delay in \snr{} for higher efficiency.
\end{myenumerateinline}

\paragraph{Summary: design implications}
%In order to address the inefficiency in \snr{}, existing and new evidences urge to \textit{systematically} treat the IO, in particular its numerous wait periods. 
%The key to address the inefficiency in \snr{} is systematically treating the \iophase{}, 
Device \snr{} shall be treated systematically.
%in particular the excessive wait periods. 
We face challenges that the invoked kernel code is diverse, complex, and cross-layer; 
%(we stress not only the driver, but all the kernel layers invoked for IO); 
we see opportunities that allow
%trading the \snr{}'s short execution delay for higher efficiency, 
focusing on hot kernel paths, 
specializing for simple concurrency, 
and gaining efficiency at the cost of increased execution time. 
\subsection{A \pmcore{} in a heterogeneous SoC}
\label{sec:bkgnd:hw}

%\input{tab-hwmodel}
% !TeX root = main.tex

%\begin{comment}
\begin{table}[t]
\centering
\scriptsize
%\footnotesize
\begin{threeparttable}
\begin{tabularx}{0.48\textwidth}{m{1.2cm}|m{0.8cm}|m{0.9cm}|m{0.9cm}|m{1.3cm}|m{1cm}}
	\hline
	\textbf{SoC} & \textbf{Cores} & \textbf{ISAs} & \textbf{Shared DRAM?} & \textbf{Mapping kern mem?} & \textbf{Shared IRQ} \\ 
	\hline   
	OMAP4460   ~\cite{omap4rtm} (2010) & A9+M3 & v7a+v7m & Full & Yes. MPU & 39/102 \\
	\hline
	AM572x~\cite{am572xtrm} (2014) & A15+M4 & v7a+v7m & Full & Yes. MPU & 32/92 \\
	\hline
	i.MX6SX~\cite{imx6sx} (2015) & A9+M4 & v7a+v7m & Full & Yes. MPU & 85/87 \\
	\hline
	i.MX7~\cite{imx7} (2017) & A7+M4 & v7a+v7m & Full & Yes. MPU & 88/90 \\
	\hline
	i.MX8M~\cite{imx8m} (2018) & A53+M4 & v8a+v7m & Full & Yes. MPU & 88/88 \\
	\hline
	MT3620~\cite{mediatek-mt3620} (2018)\tnote{*} &A7+M4 & v7a+v7m & Full & Likely. MPU & Likely most\\
	\hline
\end{tabularx}
\vspace{-1em}
	\caption{
	%	The hardware model, as the only assumptions that the \sysmodel{} builds on. Many SoCs fit the  model~\cite{omap4rtm,omap5rtm,nxp-vf6xx,exynos4120rtm}. 
	Our hardware model fits many popular SoCs which are used in popular products such as Apple Watch and Azure Sphere.
	Section~\ref{sec:eval:discussion}  discusses caveats.
%	with caveats such as not all IRQs are routed to both processors. 
	*: lack public technical details.	
}
\label{tab:hwmodel}
\end{threeparttable}
\end{table}
%\end{comment}

\paragraph{Hardware model}
%We deem that the inefficiency of device \snr{} can be greatly mitigated by offloading to a low-power, microcontroller-like core, which we refer to as a \textit{\pmcore{}}. 
%To mitigate the aforementioned inefficiency, 
We set to exploit {\pmcore{}}s already on modern SoCs. 
Hence, our software design only assumes the following hardware model which fits a number of popular SoCs as listed in Table~\ref{tab:hwmodel}.

%Many SoCs, including those in popular products (e.g., Apple Watch and Azure Sphere), largely fit the hardware model~\cite{omap4rtm,omap5rtm,nxp-vf6xx,imx6,imx7,mediatek-mt3620} as listed in Table~\ref{tab:hwmodel}. 

% https://en.wikipedia.org/wiki/Sensor_hub
% https://www.qualcomm.com/news/onq/2014/04/24/behind-sixth-sense-smartphones-snapdragon-processor-sensor-engine
% A10x
% https://en.wikipedia.org/wiki/Apple_A10X
%
% --- good but verbose --- 
%as exemplified by ARM Cortex-M3. 
%A \pmcore{} is often instantiated in conjunction with performance-oriented CPUs (e.g. ARM Cortex-A9) on modern heterogeneous SoCs. 
%Such an SoC offers energy proportionality. 
%often team up in conjunction with the performance-oriented CPU (e.g. ARM Cortex-A9) 
%
%\subsection{Architecture assumptions}
%\paragraph{Architecture characteristics}
%We make the following assumptions on the hardware architecture, and design software solely based on the assumptions. 
%We design software based on the following characteristics of a heterogeneous SoC. 
%We target heterogeneous SoCs with the following characteristics. 
\begin{myenumerate}
\item 
\textit{\textbf{Asymmetric processors}}: 
%The platform integrates the main CPU cores and a \pmcore{}. 
In different coherence domains, the CPU and the \pmcore{} offer disparate performance/efficiency tradeoffs. 
%Unlike the full-fledged CPU, 
The \pmcore{} has memory protection unit (MPU) but no MMU, incapable of running commodity OSes as-is.
%which can map at most tens memory regions.
%It cannot run commodity OSes.

\item 
\textit{\textbf{Heterogeneous, yet similar ISAs}}: 
The two processors have different ISAs, in which many instructions have similar \textit{semantics}, as will be discussed below. 
%The similarity has strong implications on our design, as will be formalized in \sect{dbt}. 
\item 
\textit{\textbf{Loose coupling}}: 
The two processors are located in separate power domains and can be turned on/off independently. 
%No cache coherence exists among core types. 
\item 
\textit{\textbf{Shared platform resources}}:
Both processors share access to platform DRAM and IO devices. 
%IO interrupts are physically wired to both processors.
Specifically, the \pmcore{}, through its MPU, should map all the kernel code/data at identical virtual addresses as the CPU does. 
	%	\note{how about MMU? linear mapping requirement}
Both processors must be able to receive interrupts from the devices of interest, e.g., MMC; 
they may, however, see different interrupt line numbers of the same device. 
\end{myenumerate}

%%\pmcore{}s are often integrated with main cores on the same SoC
%The two types of core have heterogeneous ISAs and are \textit{loosely coupled}: they have no cache coherence and reside on separate power domains. 
%\note{highlight ucontroller} The \pmcore{} resembles a microcontroller: it often lacks MMU and cannot run full-fledge OS such as Linux. 
%\note{mention shared DRAM. needed by DBT explanation}

%\paragraph{Promise of high efficiency}
\paragraph{How can \pmcore{}s save energy?}
They are known to deliver high efficiency for IO-heavy workloads~\cite{reflex,dreamweaver,turducken,somniloquy,mobilehub}. 
Specifically, they benefit the kernel's device \snr{} in the following ways.
%Specifically, it boosts the efficiency of \snr{} in two ways. 
\begin{myenumerateinline}
%\item Powering down the remainder of the SoC. 
%\item Operating autonomously.
%\item Leaving other hardware components off. 
\item 
%Autonomous operation.
%Residing in its own power domain, 
%the \pmcore{} can operate while most other SoC hardware (in particular the CPU) offline.
A \pmcore{} can operate while leaving the CPU offline.

%\item Much lower idle power. 
\item 
%Low idle power. 
%An idle \pmcore{} only consumes a few mW~\cite{k2,imx7-power}, 
%one magnitude lower than the CPU idle power. 
The idle power of a \pmcore{} is often one order of magnitude lower~\cite{k2,imx7-power},
minimizing system power during core idle periods. 

%\item Simple microarchitecture.
\item Its simple microarchitecture suits kernel execution, whose
irregular behaviors often see marginal benefits from powerful microarchitectures~\cite{mogul2008micro}.
Note that a \pmcore{} offers much higher efficiency than a LITTLE core as in ARM big.LITTLE~\cite{armbiglittle}, 
which mandates a homogeneous ISA and tight core coupling.
We will examine big.LITTLE in \sect{eval}.
% --- useful ---- 
%which mandate a homogeneous ISA and cache coherence between big and LITTLE cores.
%(they only differ in microarchitectures; weaker heterogeneity). 
%A \pmcore{}'s idle power is shown at least 10$\times$ lower than that of a LITTLE core (often tens of mW~\cite{Peters16CODES}); 
%in CPU-bound benchmarks, 
%while a \pmcore{} shows over \todo{2.5$\times$} energy efficiency compared to the CPU on the same SoC~\cite{k2,zhai17hotmobile}, 
%a LITTLE core is reported to achieve 1.1 -- 1.5$\times$ efficiency compared to a big core on the same SoC~\cite{Loghin15VLDB}. 
% --------------
%\note{workloads? shall we compare with the same BIG core?}
 
%Advanced microarchitectures that consume higher power see marginal benefit on kernel's unique execution characteristics, such as small code working set, irregular control flow, and frequent IO access

%The kernel code's unique characteristics, such as small code working set, irregular control flow, and frequent IO access, see marginal benefit from advanced microarchitectures that consume higher power~\cite{mogul2008micro}. 

%The kernel code's unique characteristics, such as small code working set, irregular control flow, and frequent IO access, see marginal benefit from advanced microarchitectures that consume higher power~\cite{mogul2008micro}. 

%Our kernel microbenchmark confirms this argument:
%in executing a CPU-intensive benchmark on a big core (Cortex-A9) and a weak core (Cortex-M3),  their respective cycle counts often differ by less than 2$\times$ while their energy consumptions differ by more than 5$\times$.
\end{myenumerateinline}

\paragraph{ISA similarity}
%On the aforementioned heterogeneous SoC,  
On an SoC we target, the CPU and the \pmcore{} have ISAs from the same family, e.g., ARM. 
%We have a key observation that many heterogeneous SoCs incorporate a CPU and a \pmcore{} with ISAs the ISA of a \pmcore{} and that of the CPU often belong to the same family. 
%Compared to the CPU, the \pmcore{} often implements a subset of instructions with same or similar \textit{semantics}, despite likely in different \textit{encoding}. 
The two ISAs often implement similar instruction \textit{semantics} despite in different \textit{encoding}. 
The common examples are SoCs integrating ARMv7a ISA and ARMv7m ISA~\cite{imx6sx,imx7,am572xtrm,mediatek-mt3620,omap4rtm}.
%Other ISA families offer their ISA options that can be integrated on the same SoC: 
Other families also provide ISAs amenable to same-SoC integration, e.g., NanoMIPS and MIPS32. % IA-32 and x86-64.
%Other ISA families offer their options for performance-oriented and efficiency-oriented ISAs that can be integrated on the same 
%is often a low-power ``profile'' (instance/member) in its ISA family.
%Compared to the performance-orient ISA in the same ISA family, 
%A modern SoC often incorporate heterogeneous cores of different ISAs of the same family.
%
We deem that the ISA similarities are \textit{by choice}.
i) For ISA designers, it is feasible to explore performance-efficiency tradeoffs within one ISA family, since the family choice is merely about instruction \textit{syntax} rather than \textit{semantics}~\cite{isa-wars}. 
%\note{cut}The designers likely start from common instruction semantics and instantiate them differently.
%they are likely to gear common instruction semantics towards different tradeoffs, 
%e.g. by customizing instruction encoding and addressing modes.
ii) For SoC vendors, incorporating same-family ISAs on one chip simplifies
%the effort in building software tools and low-level libraries~\cite{overlapping-isa},
software efforts~\cite{overlapping-isa}, silicon design, and ISA licensing. 
\subsection{OS design space exploration} 
\label{sec:bkgnd:sw}
%Given an \textit{existing} monolithic kernel, we explore the following approaches to realize heterogeneous execution.

We set to realize heterogeneous execution for 
an \textit{existing} monolithic kernel.

%We next explore software for offloading, i.e. to empower a \pmcore{} to
%execute the \iophase{} of \snr{} on behalf of the CPU, therefore leaving the CPU off.
%until the platform is fully suspended or resumed. 
%We next explore OS designs that empower a \pmcore{} to execute a commodity kernel's phases.
%This is shown in Figure~\ref{fig:offload}(b).
%A commodity kernel (e.g. Linux) cannot span over the CPU and the \pmcore{} out of box~\cite{k2,popcorn}, due to ISA heterogeneity.
%we discuss two alternative software structures and their inadequacies. 
% -- verbose --
%Our challenges are: i) complex, fast-evolving commodity kernels and ii) different ISAs and \pmcore{}'s wimpy hardware. 

%The following known techniques are inadequate.
%We discuss two popular approaches and their difficulties. 

\begin{figure}[t]
    \centering
   	\includegraphics[width=0.4\textwidth]{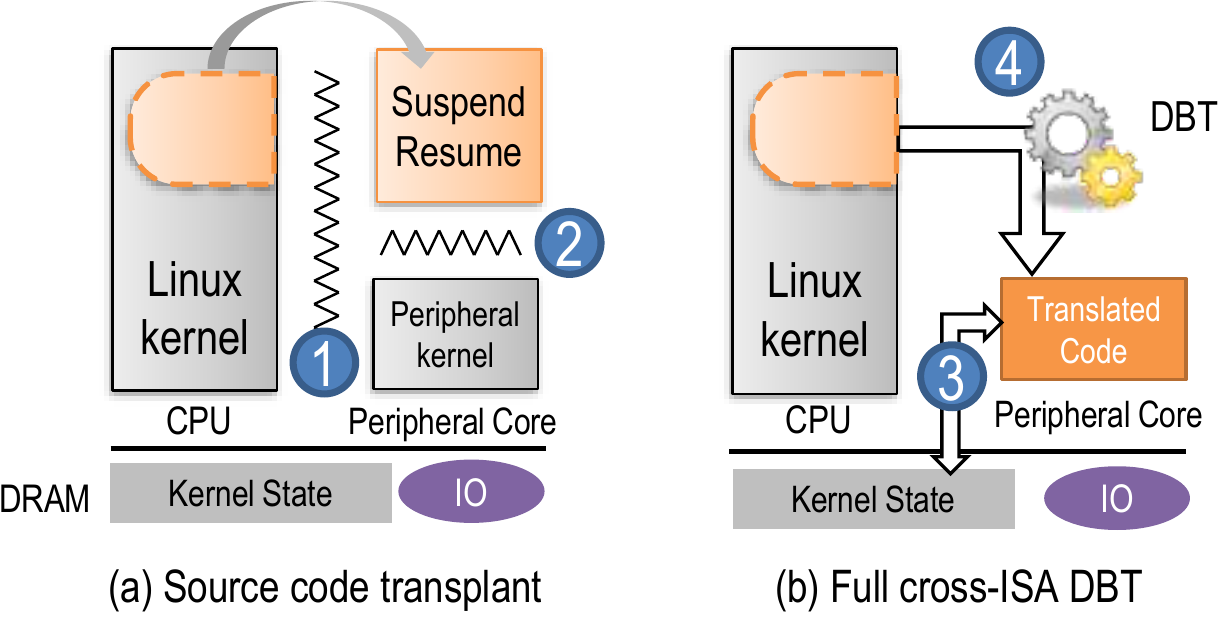}
%   	\vspace{-5pt}
    \caption{Alternative ways for offloading kernel phases}
    \label{fig:arch-cmp}
%    \vspace{-10pt}
\end{figure}

%\paragraph{How about extending existing kernel source?}
%\paragraph{How about transforming existing kernel source?}
%\note{In} presence of a monolithic kernel, we explore the following design tradeoffs.
%\paragraph{How about refactoring a monolithic kernel and cross-compiling statically?}
\paragraph{How about refactoring the kernel and cross-compiling statically?}
%\paragraph{Code transplant creates fragile interfaces}
One may be tempted to modify a monolithic kernel (we use Linux as the example below)~\cite{k2,popcorn}
%\note{Discuss modify kernel source for optmization is concurrent efforts but never enters the mainline tree.}
to be one unified source tree; 
the tree shall be cross-compiled into a kernel binary for CPU and a ``\lwk{}'' for the \pmcore{}. 
%to execute \snr{} autonomously. % for the \pmcore{}. 
%The \snr{} code runs atop the \lwk{}, which is vital to the autonomy of the offloaded execution. 
This approach results in an OS structure shown in Figure~\ref{fig:arch-cmp}(a).
%This approach results in an OS resembling a multikernel OS~\cite{multikernel}, shown in Figure~\ref{fig:arch-cmp}(a).
\begin{comment}
\note{Such} an OS differs from the existing multikernel implementation (e.g., Barrelfish OS~\cite{multikernel}) in that it relies on \textit{shared memory} instead of explicit message passing for communication.
In addition, the explicit message passing serves as an enabling factor for a multikernel OS to allow scalability and independent software evolution on different cores.
On the contrary, this approach -- operating and communicating based on shared memory in a monolithic kernel, incites many drawbacks \note{even though mapping suspend/resume to a different appears easy conceptually}.
\end{comment}
%This approach results in a multikernel OS~\cite{multikernel} shown in Figure~\ref{fig:arch-cmp}(a). 
%Though seemingly easy as it only needs to build two binaries, 
%Though seemingly easy and conceptually straightforward, 
%Though seemingly easy and conceptually \note{mapping operations to different cores}, 
%Though mapping suspend/resume to a different core may appear easy conceptually, 
Its key drawback is the two interfaces that are difficult to implement and maintain, shown as \snakezigzag{} in the figure.
%the resultant OS, however, depends on two interfaces that are difficult to implement and maintain (shown as \snakezigzag{} in the figure).

%The OS, however, depends on two interfaces that are difficult to implement and maintain (shown as \snakezigzag{} in the figure). 

%and because the wimpy \pmcore{} (a microcontroller lacking MMU) cannot afford a full Linux replica~\cite{Levasseur2004}. 

%One may be tempted to transplant the \snr{} code:
%the source is carved out from a commodity kernel, compiled for the \pmcore{}, and runs atop a lightweight kernel for the \pmcore{}. 

%This approach faces difficulties in implementing and maintaining two wide, fragile interfaces, 
%shown as \snakezigzag{} in the figure. 
%This approach results in two wide, fragile interfaces shown as \snakezigzag{} in the figure. 

% !TeX root = main.tex

%\begin{figure}[t!]
%	
%	\begin{minipage}[b]{0.15\textwidth}
%		\includegraphics[height=1.1\textwidth]{figs/plot/read}
%		        \subcaption{read}
%		\label{fig:eval:read}
%	\end{minipage}
%	\begin{minipage}[b]{0.15\textwidth}
%		%		\includegraphics[width=\textwidth]{figs/plot/scalability/wingrep.pdf}
%		\includegraphics[height=1.1\textwidth]{figs/plot/write}
%		        \subcaption{write}
%		\label{fig:eval:write}
%	\end{minipage}	
%	\begin{minipage}[b]{0.15\textwidth}
%		%		\includegraphics[width=\textwidth]{figs/plot/scalability/wingrep.pdf}
%		\includegraphics[height=1.1\textwidth]{figs/plot/fsync_2}
%		        \subcaption{fsync}
%		\label{fig:eval:fsync}
%	\end{minipage}	
%
%	%		\includegraphics[width=\textwidth]{figs/plot/scalability/wingrep.pdf}
%	\caption{A comparison of file I/O latencies
%%	\psbox{} ensures fairness by keeping throughputs of other apps largely unchanged.
%	}
%	\vspace{-15pt}		% use as needed
%	\label{fig:eval}
%\end{figure}

%\input{tab-abi}
% !TeX root = main.tex

% --- three figures, side by side --- 
\begin{figure}    
    \begin{subfigure}[t]{0.12\textwidth}
        \includegraphics[width=\textwidth]{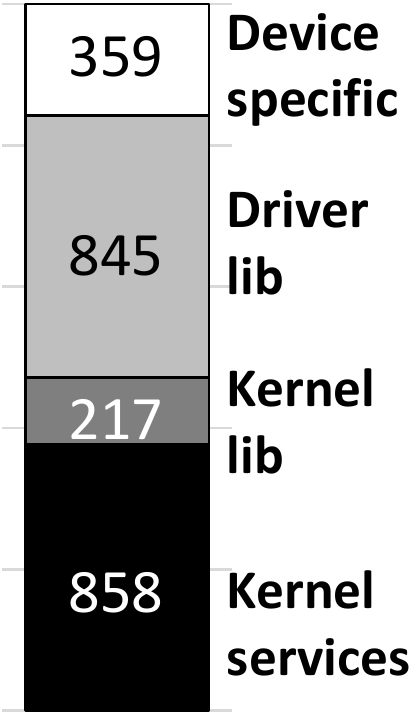}
        \caption{\# of functions}
        \label{fig:gull}
    \end{subfigure}
%    ~ %add desired spacing between images, e. g. ~, \quad, \qquad, \hfill etc. 
      %(or a blank line to force the subfigure onto a new line)
    \begin{subfigure}[t]{0.35\textwidth}
        \includegraphics[width=\textwidth]{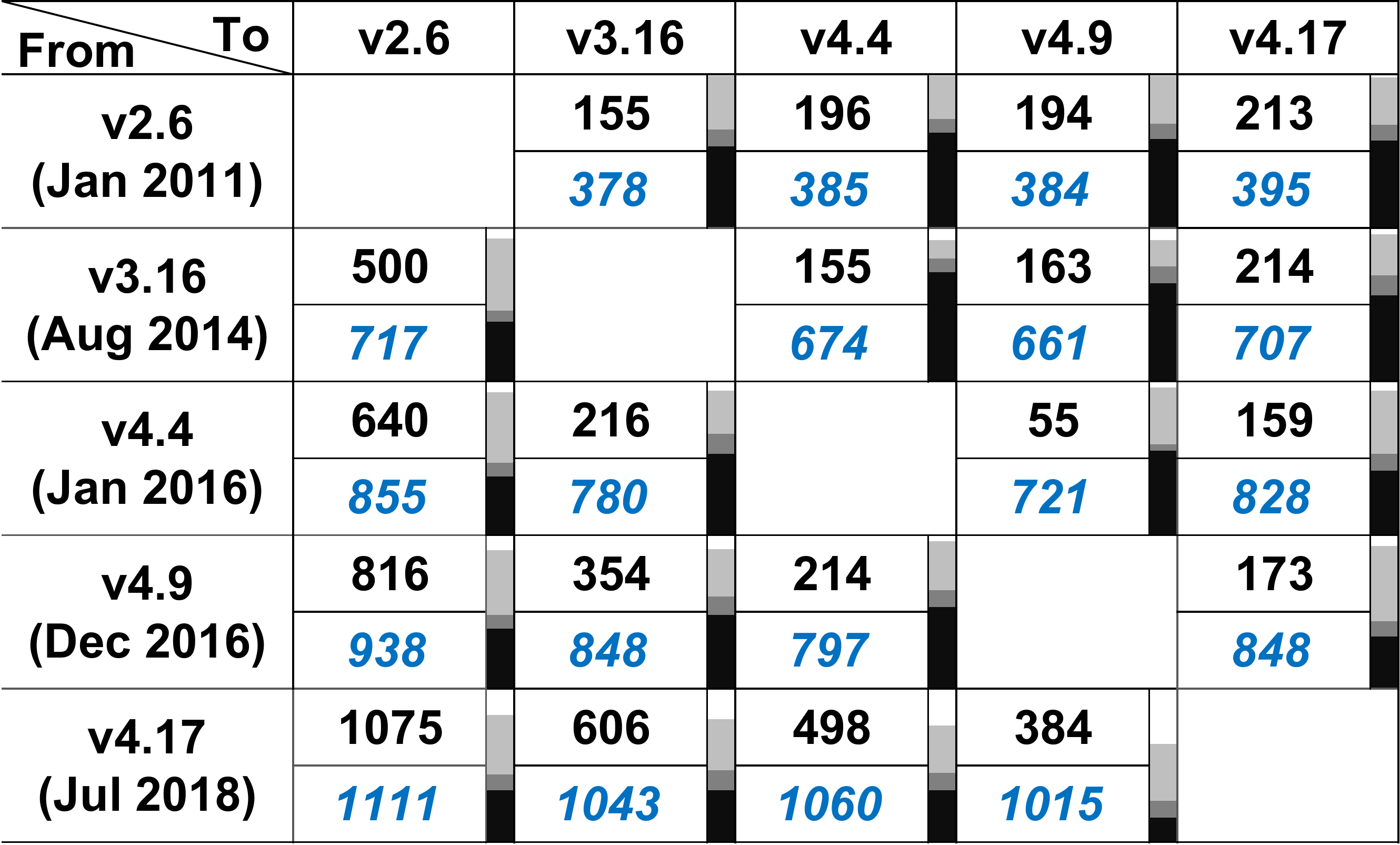}
        \caption{\# of functions (upper) \& types (lower) w/ changed ABI across kernel versions}
        \label{fig:tiger}
    \end{subfigure}
    
   	\caption{
%   	\textbf{XXX bring back the detailed figure. make the figure larger. perhaps compose with Fig 2}
   	Counts of Linux kernel functions referenced by device \snr{}, 
   	showing (a) the functions are rich and diverse
   	and (b) their ABI change is substantial over time.
   	Exported functions only. Build config: omap2defconfig. ABI changes detected with ABI compliance checker~\cite{abi-checker}}   	
   	\label{tab:abi}
%   	\vspace*{-1em}
\end{figure}

%the one between the transplanted code and the \lwk{}; 
%and the one between the two heterogeneous kernels. 
%\note{due to fast, fragile kernel interfaces (?)}

%\note{I would elaborate more. show some evidences?}
\noindent
\circled{1} The interface between two heterogeneous ISAs, as needed for resolving inter-kernel data dependency.
% for sharing the state of devices, drivers, and kernel resources.
%sharing state for device drivers, device configurations, and resource allocation. 
%the interfaces between the two heterogeneous kernels, shown as \snaketriangles{} in the figure. 
Through the interface, both kernels synchronize their kernel state, e.g., devices configurations, pending IO tasks, and locks, before and after the offloading.
%To maintain a single system image, they must synchronize shared state over this interface (e.g.  \note{an example?}) before and after the offloading.
%upon they transfer control to each other. (at handoff points)
%. \st{before suspend and after resume (before and after \snr{})}. 
%Regardless whether the synchronization is done over message passing (implemented with new messaging protocol added to kernels~\cite{helios,multikernel}) or over software shared memory (implemented as a layer underneath the kernels) ~\cite{k2,popcorn,picodriver}. 
%it is essentially built on mutual agreement on the definitions of various kernel data types, including their members and/or the memory layout. 
%Whether the interface is based on messages~\cite{helios,multikernel} or software shared memory~\cite{k2,popcorn,picodriver}, 
Built atop shared memory~\cite{k2,popcorn,picodriver},
the interface is essentially an agreement on thousands of shared Linux kernel data types, including their semantics and/or memory layout.
%Building this interface is not only tedious (for keeping the data types consistent between heterogeneous ISAs) but repetitive:
%it is essentially built on a contract on the definitions of various kernel data types, including their members and/or the memory layout. 
%This, however, essentially creates an implicit, inter-kernel ABI. 
%The ABI is highly fragile as its validity depends on the definitions of kernel data types, including their members and/or the memory layout. 
The agreement is brittle, as it is affected by ISA choices, kernel configurations, and kernel versions. 
Hence, keeping data types consistent across ISAs entails tedious tweak of kernel source and configurations~\cite{microdrivers,picodriver}.
As Greg Kroah-Hartman puts, 
``you will go insane over time if you try to support this kind
of release, I learned this the hard way a long time ago.''~\cite{stable-api-nonsense}

%Prior work showed that doing so for a single driver in a particular kernel version would take hours~\cite{microdrivers,picodriver}, let alone for many drivers in many builds multiplied by many versions. 

%Figure~\ref{tab:abi} summarizes numerous changes to the data types referenced in device \snr{}. 
%Building this interface is not only tedious (for keeping the data types consistent between heterogeneous ISAs) but repetitive: any data type change would break the interface and require to revise and rebuild the \lwk{}~\cite{microdrivers,picodriver}.
%This not only requires tedious workflow to generate the \lwk{} for , but also 
% which often require tedious workflow and even human efforts (hours)~\cite{microdrivers,picodriver}

%To implement such inter-kernel synchronization, one may install message passing interfaces to both kernels for serializing and deserializing selected kernel objects. 

\noindent
\circled{2} The interface between the transplant code and the \lwk{}, as needed for
resolving functional dependency. 
%resolving the former's functional dependency on the latter.
%the interface between the transplanted code and the \lwk{} (shown as \snakezigzag{} in the figure). 
In principle, this interface is determined by the choice of transplant boundary. 
In prior work, the example choices include the interface of device-specific code~\cite{microdrivers,picodriver,nooks},
%(which only reuses a small fraction of Linux kernel source), 
that of driver classes~\cite{sud,shadow-driver}, or that of driver libraries~\cite{k2}.
All these choices expose at least hundreds of Linux kernel functions on this interface, as summarized in Figure~\ref{tab:abi}(a). 
% thanks to diverse IO on embedded platforms (\S\ref{sec:bkgnd:code}).
This is due to Linux's diverse, sophisticated drivers.
%This is true regardless of whether the transplant code includes only the device-specific code~\cite{microdrivers,picodriver,nooks}
Implementing such an interface is daunting;
maintaining it is even more difficult due to significant ABI changes~\cite{linux-apiabi} as shown in Figure~\ref{tab:abi}(b).
%One must revise and rebuild the \lwk{} frequently to implement the updated ABI -- a moving target.
%To support transplant code ported from a different kernel release, 
%To support transplant code ported from a different kernel release, 
%the \lwk{} must be revised and rebuilt to implement the corresponding ABIs. 

% --- good. but verbose --- %
%The implication is that it is only feasible to keep the transplant boundary narrow and pick the kernel functions that have stable ABIs, as we will show in XXX. 

In summary, 
all these difficulties root in the peripheral kernel's \textit{deep dependency} on the Linux kernel. 
This is opposite to the common practice: heterogeneous cores to run their own ``firmware'' that has little dependency on the Linux kernel.
This is sustainable because the firmware stays compatible with many builds of Linux.
%These firmware hence has much higher staying power the commodity kernel.
%These firmware hence work with many builds of the commodity kernel.

\begin{comment}
\paragraph{Interfaces evolve rapidly}
% https://www.kernel.org/category/releases.html
Such interfaces are rapidly evolving and thus considered ``unstable''~\cite{linux-apiabi}. 
To show this, we investigate the signatures of exported kernel functions between recent ``longterm'' releases. 
% 3.16 -- 4.4 -- 4.9 -- 4.14
We found that overall, XX\% of exported kernel functions have changed. 
We further investigated the exported functions invoked in \snr{}. 
Of functions exported by driver libraries, XX\% API/ABI have changed. 
Of functions exported by kernel cores, XX\% ABI have changed. 
\note{what is ABI. highlight ABI is important as \sys{} acts as an external binary off the kernel tree}
\end{comment}

%\paragraph{Existing cross-ISA DBT is unaffordable to \pmcore{}s}
%\paragraph{Current cross-ISA DBT is unaffordable}
%\paragraph{How about current cross-ISA DBT?}
\paragraph{How about virtual execution?}
%DBT is a known technique to execute unmodified binary of a different ISA.  
%Recent DBTs are able to translate whole kernels. \note{cite} 
%As discussed in \sect{bkgnd}, DBT overcomes ISA gaps (and can translates kernels?).
Can we minimize the dependency?
One radical idea would be for a \pmcore{} to run the Linux kernel through virtual execution, as shown in Figure~\ref{fig:arch-cmp}(b).
%DBT is a well-known technique for executing instructions in a different ISA (called guest) on a local CPU (called . 
%A well-known technique for bridging heterogeneity ISAs, 
Powered by DBT, virtual execution allows a \textit{host} processor (e.g., the \pmcore{}) to execute instructions in a foreign \textit{guest} ISA (e.g., the CPU).
%At run time, DBT discovers guest instructions, translates them into host instructions, and executes the resultant host instructions. 
%We will discuss DBT in detail in \sect{dbt}. 
Virtual execution is free of the above interface difficulties: 
%the \pmcore{} shows the same execution behaviors as the CPU 
the translated code precisely reproduces the kernel behaviors 
and directly operates the kernel state (\circled{3}). 
%the translated execution has identical behaviors as the main kernel, directly operating on the objects of the main kernel. 
The \pmcore{} interacts with Linux through a low-level, stable interface: the CPU's ISA (\circled{4}).
%rather than the fragile ABI of Linux kernel functions. 

%However, this approach faces threefold difficulties.
%
The problem, however, is the high overhead of existing cross-ISA DBT~\cite{popcorn-app}. 
%e.g. two orders of magnitude slowdown ~\cite{popcorn-app}. 
%which is the most common reason that DBT is dismissed by systems for heterogeneous ISAs. 
%x86-ARM translation on QEMU~\cite{popcorn-app} is reported to XXX $\times$ overhead. 
%
It is further exacerbated by our \textit{inverse} DBT paradigm:
whereas existing cross-ISA DBT is engineered for % designed and optimized for
a brawny host emulating a weaker guest (e.g., an x86 desktop emulating an ARM smartphone)~\cite{mambo-x64,wang2017enabling},
our DBT host, 
%a \pmcore{} with small cache and no hardware virtualization,
a \pmcore{},
%has leaner ISA and microarchitecture (e.g. smaller cache size and lack of hardware virtualization) 
shall serve a full-fledged CPU. 
%and simpler hardware: e.g. in-order instruction pipleine, small cache size, lack of hardware virtualization support. 
%No cross-ISA DBT is optimized (ready) for this hardware architecture. 
A port of popular DBT exhibits up to 25$\times$ slowdown as will be shown in \S\ref{sec:eval}.
%The overhead would negate any benefits of \pmcore{}s and result in overall efficiency \textit{loss}. 
Such overhead would negate any efficiency promised by the hardware and result in overall efficiency \textit{loss}. 
%
%Third, cross-ISA DBT is complex~\cite{qemu}, e.g. QEMU has more than 2M SLoC. 
%Translating the Linux kernel, in particular its sophisticated core services, is complex~\cite{drk,kernel-dbt}.
%
Furthermore, cross-ISA DBT for the \textit{whole} Linux kernel is complex~\cite{qemu}. 
%QEMU~\cite{qemu} has more than 2M SLoC. 
%They are designed to run on desktop or servers (POSIX environment). 
A \pmcore{} lacks necessary environment, e.g., multiple address spaces and POSIX, for developing and debugging such complex software.
%Developing cross-ISA DBT for the \pmcore{} to translate \textit{all} Linux kernel layers, in particular the sophisticated core services~\cite{drk,kernel-dbt}, 
%is difficult, if not intractable.
%Making a \pmcore{} translate \textit{all} Linux kernel layers, in particular the sophisticated core services~\cite{drk,kernel-dbt}, 
%is difficult, if not intractable.

%\subsection{Our design objectives}
%\subsection{Design objectives for software on a \pmcore{}}
\subsection{Design objective}
\label{sec:bkgnd:obj}
%\paragraph{Design objectives}
%We seek to offload the kernel \snr{} from the main core to a \pmcore{}. 
%We execute a commodity OS (kernel) on the main core, 
%We seek to execute its \SnR{} paths on a \pmcore{}. 
%Motivated by our analysis in Section~\ref{sec:bkgnd:analysis}, we set three objectives. 
We therefore target threefold objective. 

%\item \textit{Compatibility \& Future proof?? Evolution friendly??}
%\item \textit{Kernel Reuse, Compatibility, \& Future proof}
%We recognize the tremendous efforts invested in the device drivers and their libraries in commodity kernels and set to reuse them. 
%In the face of rapid evolution of commodity kernels, we seek to reuse their device drivers and  libraries, which constitute a majority of the codebase \note{fix?}. 
%In the face of rapid evolution of commodity kernels, 
%Given the tremendous investment in commodity kernels, 
\noindent \textbf{\textit{G1. Tractable engineering.}}
We set to reuse much of the kernel source, 
%in particular their IO code that is diverse and widespread
in particular the drivers that are impractical to build anew.
%The software should have a simple structure and a lean codebase.
We target simple software for \pmcore{}s.

\noindent \textbf{\textit{G2. Build once, work with many.}}
%The software on \pmcore{} should be deployed as an optional extension to the main kernel, requiring minimal changes to the latter. 
%The software on \pmcore{} should require minimal changes to the commodity kernel, and interact with the kernel through stable interfaces. 
%In the face of kernel rapid evolution, 
%We seek good compatibility for now and future: \improve{}
One build of the \pmcore{}'s software should work with a commodity kernel's binaries built from a wide range of configurations and source versions.
This requires the former to interact with the latter through a stable, narrow ABI.  

\noindent \textbf{\textit{G3. Low overhead}}. 
%With all potential overhead, the \snr{} execution should still yield substantial efficiency gain. 
%The execution with any incurred overhead should yield substantial efficiency gain) 
The offloaded kernel phases should yield a tangible efficiency gain. 

% --- +2	3.5 page --- %
% !TeX root = main.tex

%\section{Overview}
%\label{sec:overview}

%\section{A Case for DBT}
%\section{Inverted Virtualization}
%\section{\sys{} (The XXX Model?)}
%\section{The Model of Translation atop Emulation}
\section{The \Sysmodel{} Model}
%\section{A Case for the \sysmodel{} Model}
\label{sec:case}
\label{sec:overview}

Running on a \pmcore{}, a \sysmodel{} consists of two components: 
a DBT engine for translating and executing the unmodified kernel binary;
%\st{of all device drivers, their libraries, and selected kernel services};
%TaE emulates a set of stateless kernel services.
a set of emulated, minimalistic kernel services that underpin the translated kernel code, as will be described in detail in \sect{sys}.
%The resultant \sysmodel{} binary is (shipped) deployed as a DBT engine and an emulated core, meeting our ...
A concrete \sysmodel{} implementation targets a specific commodity kernel, e.g., Linux.
A transkernel does not execute user code in ephemeral tasks as stated in \sect{intro}.

%\subsection{Software design principles}

%To achieve design objectives, 
The \sysmodel{} follows four principles:

%\paragraph{Translating the stateful code while emulating the stateless}
\paragraph{1. Translating stateful code; emulating stateless services}
%By stateful/stateless, we refer to whether specific \snr{} code share state (and therefore need to synchronize) with the kernel code outside of the \snr{} path (executed on CPU). 
%
%We translate the stateful code, including the device drivers and their libraries (e.g. sharing IO state and driver configuration) as well as a small set of kernel services, e.g. memory allocator. 
%They constitute the most diverse and widespread code invoked on the \snr{} path (\S\ref{sec:bkgnd}). 
%Through DBT, the \sysmodel{} reuses the commodity kernel code transparently without source development or build (G1). 
By \textit{stateful code}, we refer to the offloaded code that must share states with the kernel execution on CPU.
The stateful code includes device drivers, driver libraries, and a small set of kernel services. 
They cover the most diverse and widespread code in device \snr{}  (\S\ref{sec:bkgnd}). 
By translating their binaries, 
the \sysmodel{} reuses the commodity kernel without maintaining wide, brittle ABIs. (objective G1, G2)

%We choose to emulate a set of kernel services, relax their semantics to be stateless, and provide much simplified implementations. 
%These kernel services are needed by \snr{} but impose prohibitive engineering challenges to full translation on the \pmcore{}. 
%With stateless emulation, their own state only lives within the scope of one \snr{} execution, and is discarded after the execution. 

The \sysmodel{} emulates a tiny set of kernel services. 
We relax their semantics to be stateless, so that their states only live within one device \snr{} phase.
% and is discarded after the phase. 
Being stateless, the emulated services do not need to synchronize states with the kernel on CPU over ABIs.  (G2)

\paragraph{2. Identifying a narrow, stable translation/emulation ABI} 
%We determine the translation/emulation boundary to be a small set of kernel  functions. 
%We pick these functions to be classic kernel 
The ABI must be unaffected by kernel configurations and unchanged since long in the kernel evolution history. (G2)
%The narrow, stable ABI ensures a \sysmodel{}'s compatibility with 
%a specific source release of the commodity kernel under different build configurations; 
%it also significantly increases the likelihood that the build of \sysmodel{} to work with the kernel builds from different source releases (G2). 

%The narrow, stable ABI ensures that a \sysmodel{}'s compatibility is robust to the changes to the commodity kernel's build configurations.
%It also significantly increases the likelihood that the build of \sysmodel{} to work with the kernel builds from different source releases (G2). 

%This ensures the \ekc{}'s compatibility with kernel and minimizing recompilation/reengineering effort in making it work with the kernel's future releases. 
%\note{admit there's no guarantee. but the likelihood is higher}

%\paragraph{Confining offloading to only beaten paths}
%\paragraph{Focusing offloading on the beaten path and specializing for it}
%\paragraph{Focusing on the beaten path and specializing for it}
\paragraph{3. Specializing for hot paths}
In the spirit of OS specialization~\cite{exokernel,drawbridge,unikernel}, the \sysmodel{} only executes the hot path of device \snr{}; in the rare events of executing off the hot path, it transparently falls back on CPU.
%\note{whose frequency will be quantitatively shown later through our 1000-run stress test}. 
The \sysmodel{}'s emulated services seek \textit{functional equivalence} and only implement features needed by the hot path; 
they do not precisely reproduce the kernel's behaviors. (G1)
%It makes the required implementation tractable (G1). 
%It entails tractable implementation (G1).
%By doing so, we trade the complete coverage of offloading and precise kernel behaviors for engineering ease, making the implementation task tractable (G1). 

%\item
%\paragraph{Emulation: Specializing the implementations}
%Rather than precisely preserving of the main kernel behaviors, 
%\ekc{} seeks \textit{functional equivalence}: \ekc{} only implements the semantics needed by \snr{} -- under the same kernel ABI.

\paragraph{4. Exploiting ISA similarities for DBT}
The \sysmodel{} departs from generic cross-DBT that bridges arbitrary guest/host pairs; 
it instead systematically exploits similarities in instructions semantics, register usage, and control flow transfer.
%\st{We bypass the intermediate representation commonly seen in cross-ISA DBT, and directly reuse instruction selection, register allocation, and control flow transfer.}
This makes cross-ISA DBT affordable. (G3)
%This makes cross-ISA DBT affordable to the \pmcore{} and ultimately makes \sysmodel{} practical.

%\subsection{Hardware assumptions}
%\note{should go to S2.2. This section is about our design already}	
%%\note{summarize what the hardware should do for Transkernel to work}
%In addition to an architectural overview of a peripheral core in a heterogeneous SoC provided by section~\ref{sec:bkgnd:hw}, we summarize and restate the detailed hardware model required by \sysmodel{} as follows.
%
%Transkernel runs on a heterogeneous SoC, which has \textbf{asymmetric cores}.
%These cores run \textbf{heterogeneous yet similar ISAs}, where the ISA on the peripheral core is (presumably) \textit{simpler} but has similar semantics with the ISA on CPU;
%these cores are \textbf{loosely coupled}, which can be turned on and off independently.

\paragraph{Limitations}
First, across ISAs of which instruction semantics are substantially different, e.g., ARM and x86, the \sysmodel{} may see diminishing or even no benefit. 
Second, the \sysmodel{}'s longer delays (albeit lower energy) may misfit latency-sensitive contexts, e.g., for waking up platforms in response to user input. 
Our current prototype relies on heuristics to recognize such contexts and falls back on the CPU accordingly (\sect{sys}).

In \sect{sys} below we describe how to apply the model to a concrete \sysmodel{}, in particular our translation/emulation decisions for major kernel services, and our choices of the emulation interface.
We will describe DBT in \sect{dbt}.

\begin{figure}[t]
    \centering
   	\includegraphics[width=0.4\textwidth]{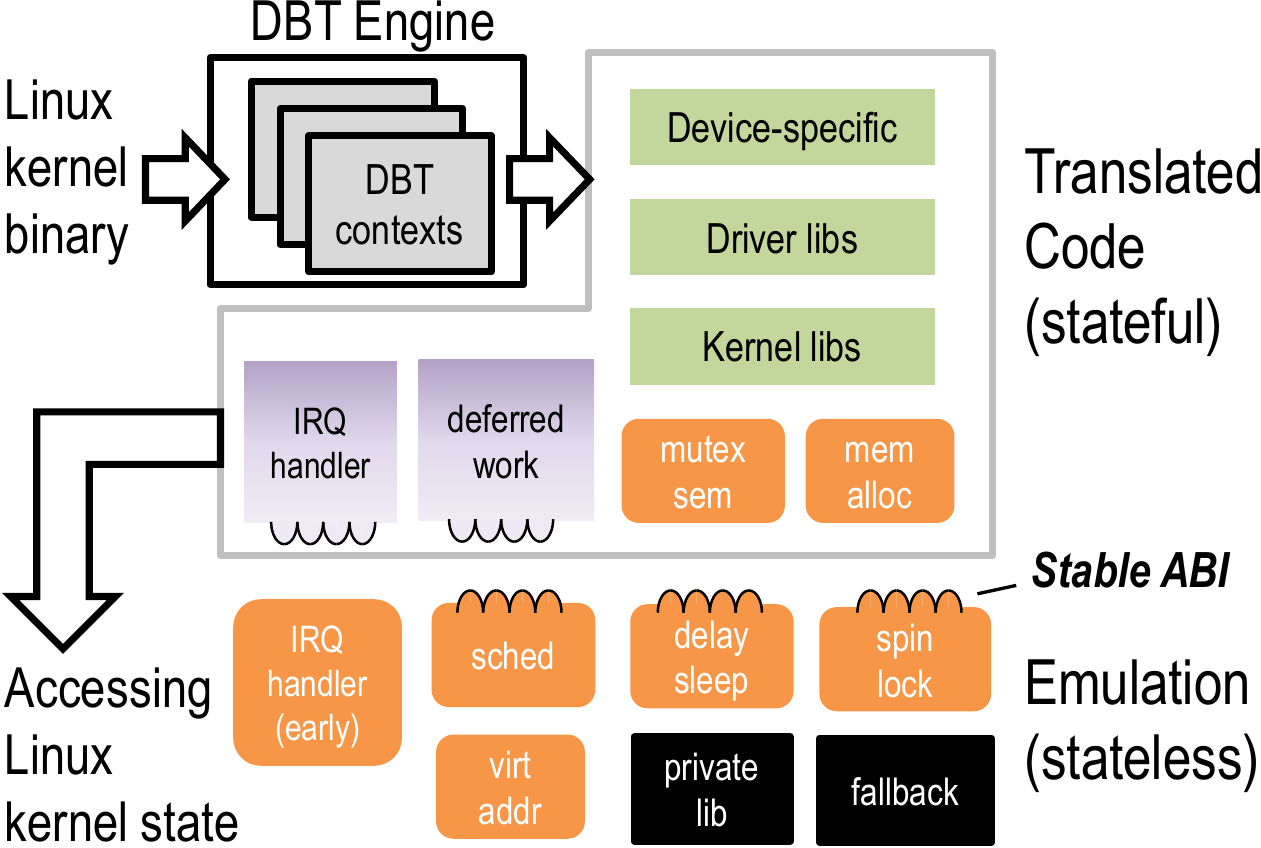}
    \caption{The \sys{} structure on a \pmcore{}}
    \label{fig:arch}
    \vspace{-10pt}
\end{figure}

% --- +.5	4 page --- %
% !TeX root = main.tex

% \section{Emulation of Selected Kernel API}
% \section{Emulation of Kernel Infrastructure}
% \section{The Emulated Kernel Core}
%\section{\sys{}}
%\section{ARK: A \Sysmodel{} Implementation}
\section{ARK: An ARM Transkernel}
\label{sec:sys}
%Ideally, the design of \sys{} should be agnostic to the kernel design, so that it can be ``future proof'', i.e. working with future kernel releases without being redesigned or even recompiled.
%This however requires \sys{} to honestly translate most, if not all the kernel and incurs almost intractable engineering effort on a \pmcore{}. 
%%We hereby make the following trade-off, by make \sys{} exploits \textit{stable} features of the kernel, such as mature (?) kernel APIs and memory layout invariants. \note{cite?}
%We hereby make the following trade-off, by building in \sys{} the knowledge of \textit{stable} features of the kernel, such as mature (?) kernel APIs and memory layout invariants. 
%Such knowledge may not be permanently invariant. 
%However, its evolution is relatively slow. And re-engineering the translator will be much easier than pushing new changes to the kernel.

%\paragraph{Overview}
Targeting an ARM SoC, we implement a \sysmodel{} called \sys{}.
%\sys{} targets an ARM SoC fitting our hardware model in \sect{bkgnd:hw}. 
The SoC encompasses a popular combination of ISAs: 
ARMv7A for its CPU and ARMv7m for its \pmcore{}.
The CPU runs Linux v4.4. 
%\sect{eval} describe other hardware details of the SoC. 
%on which Cortex-A9 (the CPU) and Cortex-M3 (the \pmcore{}) share access to DRAM and IO (see \sect{eval} for hardware details). 
%The kernel running on the CPU is Linux (v4.4). 

%\paragraph{The \snr{} workflow}
\paragraph{Offloading workflow}
\label{sec:emu:workflow}
\sys{} is shipped as a standalone binary for the \pmcore{},
accompanied by a small Linux kernel module for control transfer between CPU and the \pmcore{}.
We refer to such control transfer as \textit{handoff}.
%between the CPU and the \pmcore{}s.
%\sys{} is an add-on to the kernel; without \sys{}, the OS operates (\snr{}) as before in the traditional way. 
%To start \sys{}, the Linux kernel loads the \sys{} binary for the \pmcore{} and turns on (executes) switches to/from the \pmcore{} during \snr{}. 
%To load \sys{}, the kernel loads the executable binary to the shared memory and instructs the \pmcore{} to execute the binary; 
Prior to a device suspend phase, the kernel 
shuts down all but one CPU cores, passes control to the \pmcore{}, and shuts down the last CPU core.
%In the early stage of a suspend procedure, the kernel transfers control to \sys{} prior to invoking any IO code: it shuts down all non-boot cores, transfers the control to the \pmcore{}, and shuts down the main core immediately. \note{has implication on spinlock design}
%\sys{} uses one MPU entry for each IO region and time-multiplexes the limited MPU entries among IO regions. 
%\sys{} does not support the offloaded execution to modify the kernel address space, e.g. by calling ioremap() or kmap(), treating it off the beaten path. 
%We have not observed that \snr{} modifies existing mappings. 
Then, \sys{} completes the \iophase{} in order to suspend the entire platform. 
Device resume is normally executed by \sys{} on the \pmcore{}; in case of urgent wakeup events (e.g., a user unlocking a smart watch screen), the kernel resumes on CPU with native execution.

\paragraph{System structure}
As shown in Figure~\ref{fig:arch}, \sys{} runs a DBT engine, its emulated kernel services, and a small library for managing the \pmcore{}'s private hardware, e.g., interrupt controllers. 
%According to the \sysmodel{} model, 
%\sys{} translates all device-specific code, the libraries invoked by them, and a few kernel services we deem must be stateful (summarized in Table~\ref{tab:services} and examined below). 
%Its \ekc{} provides stateless emulation for the remaining kernel services. 
%The interface between the translated code and the \ekc{} are a small set of functions with stable ABI: 
%The translated code interacts with the emulated services through a narrow, stable ABI: %summarized in Table~\ref{tab:ker-api}.
%a small set of functions with stable ABI: 
The emulated services serves downcalls (\snakebump{}) from the translated code and makes upcalls (\invsnakebump{}) into the translated code.
%The interface is illustrated as  in Figure~\ref{fig:arch} and summarized in Table~\ref{tab:ker-api}.
Table~\ref{tab:ker-api} summarizes the interfaces. 
%The interfaces of downcall and upcall are narrow; they are stable ABI in the existing Linux kernel as summarized in Table~\ref{tab:ker-api}. 
%Table~\ref{tab:ker-api} summarizes the emulated services and their interfaces.
%We will describe the details on intercepting function calls across the emulation boundary in Section~\ref{sec:impl}. 
Upon booting, \sys{} replicates Linux kernel's linear memory mappings for addressing kernel objects in shared memory~\cite{k2,picodriver}.
\sys{} maps I/O regions with MPU and time-multiplexes the regions on the MPU entries.

%Under the DBT, it provides a runtime that emulates a set of kernel services (a virtual address space?) and implements low-level routines for managing \pmcore{}'s private resources, e.g. cache controller and interrupt controller, which are absent in the main kernel. \note{improve}
%\ekc{} is stateless (except for \snr{} abort described in \sect{abort}). 
%\ekc{} exports functions to be invoked by the translated code (downcall) and invokes the translated code (upcall).
%Both downcall and upcall interfaces are stable ABI in the existing Linux kernel. 
%We will describe the details on intercepting function calls across the emulation boundary in Section~\ref{sec:impl}. 
%Table~\ref{tab:ker-api} summarizes the emulated services and their interfaces.

To support concurrency in the offloaded kernel phases, \sys{} runs multiple DBT contexts. 
Each context has its own DBT state (e.g., virtual CPU registers and a stack), executing DBT and \ekc{} independently.
Context switch is as cheap as updating the pointer to the DBT state.
%\sys{} switches among multiple dispatcher contexts in a cooperative manner. 
%During \snr{}, all interrupts are routed to the \pmcore{}, prompting \sys{} to translate the guest kernel's interrupt handlers. 

\sys{} executes the hot paths.
Upon entering cold branches pre-defined by us, e.g., kernel \code{WARN()},
\sys{} migrates all the DBT contexts of \textit{translated} code back to the CPU and continues as \textit{native} execution there (\S\ref{sec:abort}).

%\snr{} path, irq handlers, and deferred kernel work (as described in \sect{bkgnd}). 
%Each dispatcher context has its own emulated registers for DBT and stack, and independently executes DBT (translated code) and the \ekc{}. 
%\sys{} switches among multiple dispatcher contexts in a non-preemptive manner. 

%ECKLib interfaces with the translated code by exporting functions to be invoked by the latter (downcall) and by calling functions of the latter (upcall). 
%EKCLib exposes interfaces to be invoked by the translated code (downcall) and will also invoke the translated code (upcall); 
%a library of functions that are ABI-compatible with their counterparts in the kernel. 
%these functions will be invoked by the translated kernel code (downcall) and they may invoke the translated kernel code (upcall); 

%We will describe the details on intercepting and resolving function calls across the emulation boundary in Section~\ref{sec:impl}. 

%\input{tab-ker-api}

% !TeX root = main.tex

\begin{table}[t]
    \centering

	\footnotesize
	\begin{subfigure}[t]{\linewidth}
    \begin{tabular}{ll}

    \toprule[0.7pt]%\midrule[0.3pt]
    \textsf{\textrm{Kernel services}} & \textsf{\textrm{Implementations \& reasons}} \\ 
    \toprule
		Scheduler (\S\ref{sec:emu:sched}) & Emulated. Reason: simple concurrency. \\
%		Virt addr space (\S\ref{sec:emu:addr}) & Emulated. Abort on modifications.\\
		IRQ handler (\S\ref{sec:emu:irq}) & Early stage emulated; then translated \\
		HW IRQ controller (\S\ref{sec:emu:irq}) & Emulated. Reason: core-specific \\
		Deferred work 	(\S\ref{sec:emu:tasklet}) & Translated. Reason: stateful \\
		Spinlocks (\S\ref{sec:emu:lock}) & Emulated. Reason: core-specific \\
%		Sleepable locks (\S\ref{sec:emu:lock}) & Fast path translated. Must be stateful (slow path calls into emulated services) \\
		Sleepable locks (\S\ref{sec:emu:lock}) & Fast path translated. Reason: stateful\\
		Slab/Buddy allocator (\S\ref{sec:emu:mm}) & Fast path translated. Reason: stateful\\
		Delay/wait/jiffies (\S\ref{sec:emu:time}) & Emulated. Reason: core-specific \\
%		Update to jiffies (\S\ref{sec:emu:time}) & E \\
    \midrule[0.7pt]
    %\midrule[0.3pt]\bottomrule[1pt]
    \end{tabular}
    %\caption{Major kernel services supported by \sys{}}
%        \label{tab:services}
	\end{subfigure}

    \begin{subfigure}[t]{\linewidth}
        \centering
    
    	\footnotesize
    
        \begin{tabular}{c}
    	\toprule[0.7pt]
    		 jiffies \hfill \hspace{0.42cm} udelay() \hfill \hspace{0.42cm} msleep() \hfill \hspace{0.42cm} tasklet\_schedule() \hfill \hspace{0.42cm} irq\_thread()\\
    		ktime\_get() \hfill queue\_work\_on() \hfill worker\_thread() \hfill run\_local\_timers() \\
    		generic\_handle\_irq() \hfill schedule() \hfill async\_schedule()* \hfill  do\_softirq()*\\
        	\midrule[0.7pt]
        	%\midrule[0.3pt]\bottomrule[1pt]
        \end{tabular}

    	*=ABI unchanged since 2014 (v3.16); others unchanged since 2011 (v2.6).
        %\caption{A Summary of all interface that \sys{} relies on}
    %    \vspace{-10pt}
    \end{subfigure}

%   	\caption{Major Kernel services(top) and API(bottom) supported by \sys{}}
   	\caption{Top: Kernel services supported by \sys{}. Bottom: Linux kernel ABI (12 funcs+1 var) \sys{} depends on.
   	\sys{} offers complete support for device \snr{} in Linux.
   	}
    \label{tab:services}
    \label{tab:ker-api}
    \vspace*{-2.0em}
\end{table}

%\subsection{Concurrency model}
%\note{Should this be brought upfront? to overview?}
%
%The only (?) scheduling function it requests is for code asynchronous execution through workqueue.
%%, e.g. to schedule a future check of a device state, for which it invokes XXXX (tasklet functions). 
%For instance, the \snr{} path iterates through drivers, requesting all device-specific callbacks to be executed asynchronously, and executes them later. 
%
%The emulated kernel core greatly simplified the multicore, preemptive concurrency presented by the kernel to be a single-core, non-preemptive executor that switches among three contexts:
%it executes the \snr{} path in one main thread \note{bad term -- function context? corecall context?}; 
%upon interrupts, it switches to execute the interrupt handlers (will be discussed later); 
%it executes the asynchronous work items (submitted by the main thread and interrupts) whenever the main thread goes to sleep or waits for the work item completion. 
%\note{For each context}, \sys{} may execute in the EKC or codecache. 

%\subsection{Multiple dispatcher contexts}
%\subsection{Scheduling Multiple Dispatcher Contexts}
%\subsection{Scheduling Multiple Dispatcher Contexts}
\subsection{A Scheduler of DBT Contexts}
\label{sec:sys:sched}
\label{sec:emu:sched}

%\input{fig-context}

%\sys{} maintains multiple dispatcher contexts and voluntarily switch among them to emulate the 
%concurrent kernel executions of the syscall path, the interrupt handlers, and deferred kernel tasks .

\sys{} emulates a scheduler which shares no state, e.g., scheduling priorities or  statistics, with the Linux scheduler on the CPU. 
Corresponding to the simple concurrency model of \snr{} (\S\ref{sec:bkgnd}),
\sys{} eschews reproducing Linux's preemptive multithreading but instead maintains and switches among cooperative DBT contexts: 
%one primary context for executing the\note{hot} path of \snr{},
one primary context for executing the syscall path of \snr{},
one for executing IRQ handlers (\S\ref{sec:emu:irq}),
and multiple for deferred work (\S\ref{sec:emu:tasklet}). 
Managing no more than tens of contexts, \sys{} uses simple, round-robin scheduling. 
It begins the execution in the syscall context; 
when the syscall context blocks (e.g., by calling \code{msleep()}), \sys{} switches to the next ready context to execute deferred functions until they finish or block. 
When an interrupt occurs, \sys{} switches to the IRQ context to execute the kernel interrupt handler (\S\ref{sec:emu:irq}).
%When a hardware interrupt occurs, \sys{} switches to the IRQ context to execute the kernel interrupt handler (\S\ref{sec:emu:irq}).

%\input{emu-addr}

% !TeX root = main.tex

\subsection{Interrupt and Exception Handling}
\label{sec:emu:irq}

During the offloaded \iophase{}, all interrupts are routed to the \pmcore{} and handled by \sys{}. 
\paragraph{Kernel interrupt handlers}
%\sys{} translates most portion of the guest kernel's handlers. 
\sys{} emulates a short, early stage of interrupt handling 
while translating the kernel code for the remainder. 
This is because this early stage is ISA-specific (e.g., for manipulating the interrupt stack), on which the CPU (v7a) and the \pmcore{} (v7m) differ. 
% is specific to the main CPU's ISA (v7a), which mismatches the \pmcore{} (v7m), e.g. due to different formats of IRQ stack frames. 
Hence, the \ekc{} implement a v7m-specific routine and install it as the hardware interrupt handler. 
Once an interrupt happens, the routine is invoked to
finish the v7m-specific task and make an upcall to the kernel's ISA-neutral interrupt handling routine (listed in Table~\ref{tab:ker-api}),
%This prevents \sys{} from directly translating/executing the guest code. 
%The EKC therefore provides such low-level interrupt routines as part of its emulation library and makes an upcall to the XXX function (listed in table XXX), which is the entrance to the hardware-independent interrupt handling of the guest kernel. 
from where the \sys{} translates the kernel to finish handling the interrupt. 

\paragraph{Hardware interrupt controller} 
\sys{} emulates the CPU's hardware interrupt controller. 
This is needed as the two cores have separate, heterogeneous interrupt controllers.
%Therefore, the kernel code that manipulates the CPU's controller, e.g. 
%for masking interrupt sources or setting interrupt priorities,
%does not operate on the \pmcore{}'s controller. 
%\sys{} emulates the interrupt controller at the level of controller registers. 
The CPU controller's registers are unmapped in the \pmcore{}; upon accessing them (e.g., for masking interrupt sources) the translated code triggers faults. 
\sys{} handles the faults and operates the \pmcore{}'s controller accordingly.
%\note{We will describe more implementation details in} \sect{eval:discussion}.

%Since the addresses of these registers are unmapped in the \pmcore{}'s MPU,
%the translated execution triggers hardware faults upon accessing these registers (e.g. for masking interrupt sources or setting interrupt priorities). 
%\sys{} handles the faults; 
%based on the faulty addresses and instructions, it operates on the \pmcore{}'s interrupt controller accordingly.  
%EKC handles the faults, examines the faulty address (IO registers) and instruction, and convert the register reads/writes to  operations on m3's NVIC. 
 
\paragraph{Exception: unsupported}
We don't expect any exception in the offloaded kernel phases. 
In case exception happens, \sys{} uses its fallback mechanism (\S\ref{sec:abort}) to migrate back to CPU. 
%as kernel subsystem (e.g., device drivers, kernel services, etc.) on our beaten path are always mapped. If this happens, we use the migration mechanism to fall back to the CPU. %\note{Liwei - fix}
%https://stackoverflow.com/questions/8345300/can-vmalloc-pages-be-swapping-pages

% !TeX root = main.tex

%\subsection{Preserving Asynchronous Semantics}
%\subsection{Deferred execution -- scheduling?}
%\subsection{Workqueues}
\subsection{Deferred Work}
\label{sec:emu:tasklet}
%EKC emulates two most common facilities for deferred execution: threaded IRQ and workqueue. 
%EKC emulates the following common facilities.

%The Linux kernel provides services for deferring executing work as functions. 
Device drivers frequently schedule functions to be executed in the future. 
\sys{} translates the Linux services that schedule the deferred work
as well as the actual execution of the deferred work.
%it also translates the actual execution of the deferred work. 
%The translation choice is because such services must be \textit{stateful}:
ARK chooses to translate such services because they must be \textit{stateful}:
the \pmcore{} may need to execute deferred work created on the CPU prior to the offloading, e.g., freeing pending WiFi packets; 
it may defer new work until after the completion of resume. 

\sys{} maintains dedicated DBT contexts for executing the deferred work (\sect{sys:sched}).  
While the Linux kernel often executes deferred work in kernel threads (daemons), our insight is that deferred work is oblivious to its execution context (e.g., a real Linux thread or a DBT context in \sys{}). 
Beyond this, \sys{} only has to run the deferred work that may \textit{sleep} with separate DBT contexts so that they do not block other deferred work.
From these DBT contexts, \sys{} translates the main functions of 
the aforementioned kernel daemons, which retrieve and invoke the deferred work. 

%For each deferred execution facility supported, we describe its execution context and its emulated interface. 

%\paragraph{Threaded IRQ} 
%A popular facility since 2.6.30 (?).
%A standard facility for IRQ bottom-halves \note{cite}
\begin{comment} % --- too verbose --- %
\noindent
\textbf{Threaded IRQ} is 
a popular kernel service for executing the \textbf{static} heavy-lifting work in IRQ handling (``bottom-halves''). 
It allows bottom-halves to run in a thread context rather than the hardware interrupt handler. 
When registering (initializing?) an IO device, the kernel registers a function called threaded IRQ handler for each IRQ source;
once an IRQ fires, the service schedules the function to be executed in a kernel thread. 
As kernel allows threaded IRQ handlers to sleep,
%To prevent them from blocking other deferred work, 
\sys{} maintains per-IRQ contexts for executing deferred threaded IRQ functions. 
%one for each IRQ source. . 
Each context makes upcalls into \_\_irq\_wake\_thread() (listed in Table~\ref{tab:ker-api}) for examining and executing their corresponding threaded IRQ handlers and yield when they block. 
\end{comment}

\noindent
\textbf{Threaded IRQ} defers heavy-lifting IRQ work (i.e., deferred work) to a kernel thread which executes the work after the hardware IRQ is handled. 
%For each IRQ source, driver registers one threaded IRQ handler. 
A threaded IRQ handler may sleep. 
Therefore, \sys{} maintains per-IRQ DBT contexts for executing these handlers. 
%one for each IRQ source. . 
Each context makes upcalls into 
\code{irq\_thread()} (the main function of threaded irq daemon, listed in Table~\ref{tab:ker-api}). 
%for examining and executing their corresponding threaded IRQ handlers. 

\begin{comment}
EKC emulates the scheduler interface for submitting the (\_\_irq\_wake\_thread, ABI unchanged since XXX). 
\sz{The abi is not compatible even between 4.4 to 4.17; however, emulation only relies on second argument struct irqaction, which is stable since 4.4(Jan 2016)}
When new thread functions are submitted, EKC queues them for respective thread context. 
Later, when context scheduler switches to the context it executes queued functions in order. 
In case a function blocks, the context scheduler switches to the next context as have been described in XXX. 
\end{comment}

%\paragraph{Tasklet \& timer callbacks}
%\paragraph{Tasklets \& workitems}  
\paragraph{Tasklets, workitems, and timer callbacks}  
%\lwg{The problem of tasklet-schedule is how to deal with residue states (e.g., daemon threads) -- a generic way is to 1) identify and 2)translate worker function , which is do-softirq in this case. The rationale behind this is tasks/work items are statically put and only waiting to be executed (i.e., using call back). This is exemplified by do-softirq.}
The kernel code may dynamically submit short, non-sleepable functions (tasklets) or long, sleepable functions (workitems) for deferred execution. 
%Specifically, each workitem is submitted to a specific workqueue, where all workitems will be executed in the queue order. 
Kernel daemons (softirq and kworker) execute tasklets and workitems, respectively.

%Since tasklets are submitted and handled centrally while workqueues are separate, 
\sys{} creates one dedicated context for executing all non-sleepable tasklets and per-workqeueue contexts for executing workitems so that one workqueue will not block others. 
These contexts make upcalls to the main functions of the kernel daemons (\code{do\_softirq()}, \code{worker\_thread()}, and \code{run\_local\_timers()}), translating them for retrieving and executing deferred work. 
\subsection{Locking}
\label{sec:emu:lock}

%\paragraph{Spinlocks}
\textbf{Spinlocks} %are for protecting short critical sections.
\sys{} emulates spinlocks, because their implementation is core-specific and that
%All spinlocks are stateless with regard to one execution of \snr{}: 
\sys{} can safely assume all spinlocks are free at handoff points:
%Their states never have to be synchronized between kernels, 
as described in early \sect{sys}, handoff happens between one CPU core and one \pmcore{}, which do not hold any spinlock; 
all other CPU cores are offline and cannot hold spinlocks. 
%Because of this, \sys{} emulates spinlocks as their interfaces have stable ABI and that their locked/free states are never shared between \sys{} and the main kernel. \note{improve}
Hence, \sys{} emulates spinlock acquire/release by pausing/resuming interrupt handling. 
This is because \sys{} runs on one \pmcore{} and the only hardware concurrency comes from interrupts.
%interrupt handlers are translated as described in \sect{emu:irq}, 
%this function simply disables (DBT) switches to the dispatcher context for interrupt handling. \note{check this}
%this function simply instructs the scheduler pauses switches to the context for interrupt handling. 
%Unlock is just resuming such switches. 
%toggles off/on interrupts on the \pmcore{} to disable concurrency between interrupts 

\paragraph{Sleepable locks} 
\sys{} translates sleepable locks (e.g., mutex, semaphore) 
because these locks are stateful: 
for example, the kernel's clock framework may hold a mutex preventing \snr{} from concurrently changing clock configuration~\cite{linux-clk}.
Furthermore, mutex's seemingly simple interface (i.e., compare \& exchange in fast path) has \textit{unstable} ABI and therefore unsuitable for emulation:
a mutex's reference count type changes from \code{int} to \code{long} (v4.10), breaking the ABI compatibility. 
The translated operations on sleepable locks may invoke spinlocks or the scheduler, 
e.g., when updating reference counts or putting the caller to sleep, 
%e.g. when it fails to acquire the lock and therefore needs to update reference count and puts the caller to sleep. 
for which the translated execution makes downcalls to the \ekc{}.% as described above. 
In practice, no sleepable lock is held prior to system suspend.

\subsection{Memory Allocation}
\label{sec:emu:mm}

The \iophase{} frequently requests dynamic memory, often at granularities of tens to hundreds of bytes.
% via \code{kmalloc/kzalloc()}. 
By Linux design, such requests are served by the kernel slab allocator backed by a buddy system for page allocation (fast path); 
when the physical pages runs low, the kernel may trigger swapping or kill user processes (slow path).
%https://events.static.linuxfound.org/images/stories/pdf/klf2012_kim.pdf

\sys{} provides memory allocation as a stateful service. 
It translates the kernel code for the fast path, including the slab allocator and the buddy system. 
%In the rare case that the allocation enters the slow path (e.g. due to low physical memory), \sys{} aborts offloading. 
In the case that the allocation enters the slow path (e.g., due to low physical memory), \sys{} aborts offloading;
fortunately, our stress test suggests such cases to be extremely rare, as will be reported in \sect{eval}.
With a stateful allocator, the offloaded execution can free dynamic memory allocated during the kernel execution on CPU, 
and vice versa. 
Compare to prior work that instantiates per-kernel allocators with split physical memory~\cite{k2}, \sys{} reduces memory fragmentation and avoids 
%tracking the ownership of each dynamic memory allocation. 
tracking \textit{which} processor should free \textit{what} dynamic memory pieces. 
Our experience in \sect{eval} show that \sys{} is able to handle intensive memory allocation/free requests such as in loading firmware to a WiFi NIC.

\subsection{Delays \& Timekeeping}
\label{sec:emu:time}
%\subsection{Timekeeping}

\paragraph{Delays}
\sys{} emulates \code{udelay()} and \code{msleep()} for busy waiting and sleeping. % which are widely used by \snr{}. 
\sys{} converts the expected wait time to the hardware timer cycles on the \pmcore{}. 
\sys{} implements \code{msleep()} by pausing scheduling the caller context. 
%For \code{msleep()}, \sys{} switches from the current context (sleeping) $T$ to the next ready one and refrains from switching to $T$ until the expected time has elapsed. 
%EKC emulates \code{msleep()} ... with \pmcore{} clock ... converting clockrate... 

%\paragraph{Time keeping}
%\paragraph{\code{jiffies}}
\paragraph{jiffies}
The Linux kernel periodically updates jiffies, a global integer, as a low-overhead measure of elapsed time. 
%\sys{} emulates the updates to jiffies by consulting the \pmcore{}'s hardware timer. 
By consulting the \pmcore{}'s hardware timer, \sys{} directly updates the jiffies.
%\sys{} directly accesses the jiffies variable 
%jiffies is thus the only shared variable on \sys{}'s emulated interfaces (others are all functions).
It is thus the only shared variable on the kernel ABI that \sys{} depends (all others are functions).

\section{The Cross-ISA DBT Engine}
\label{sec:dbt}
%\vspace{-10pt}

%In summary, to achieve our goals with the DBT approach, we need unconventional DBT designs that are powerful enough for one order of magnitude performance boost while simple enough to be engineered for a \pmcore{}. 

%The success of \sys{} hinges on its DBT performance. 
%Yet, the overhead of current cross-ISA DBT is one order of magnitude higher than what is needed for \sys{} to gain energy efficiency. 

% !TeX root = main.tex
%\subsection{A Primer on Dynamic Binary Translator}

%\subsection{A Cross-ISA DBT Primer}

%\paragraph{Full kernel DBT is unaffordable}
%\paragraph{Kernel-level, cross-ISA DBT is complex and expensive}
%\paragraph{Existing DBT is unaffordable for \pmcore{}s}
%DBT is a known technique to execute unmodified binary compiled in ISA X on a CPU with ISA Y 

\paragraph{A Cross-ISA DBT Primer}
%DBT is a known virtualization technique~\note{cite}, 
DBT, among its other uses~\cite{valgrind,pin,dynamorio}, is a known technique allowing a \textit{host} processor to execute instructions in a foreign \textit{guest} ISA.
%\paragraph{Dynamic code discovery and translation}
In such cross-ISA DBT, 
%a program, called DBT engine, runs on a \textit{host} CPU; 
the host processor runs a program called DBT engine.
At run time, 
the engine reads in guest instructions, translates them to host instructions based on the engine's built-in \textit{translation rules}, and executes these host instructions. 
%the DBT engine dynamically discovers and translates guest instructions in unit of translation block. 
%the DBT engine dynamically discovers a sequence of guest instructions, translates them to host instructions, and  in unit of translation block. 
The engine translates guest instructions in the unit of translation block -- a sequence (typically tens) of guest instructions that has one entry and one or more exits. 
%A typical block consists of tens of instructions. 
%Execution always starts at the beginning of TBs, and may not necessarily finish entire block depending on instructions and CPU flags status.
%The boundary of each TB may vary depending on DBT's choice, but the common practice is whenever it hits a jump(e.g., branch or function call), the current TB should be closed.
After translating a block, the engine saves the resultant host instructions to its \textit{code cache} in the host memory, so that future execution of this translated block can be directed to the code cache.
%For more background on generic DBT, see prior work~\cite{drk}.
%\st{The DBT engine typically \textit{chains} host code blocks for performance: 
%it installs instructions into the code blocks in the code cache, so that these blocks can directly branch to each other without invoking the DBT engine.}

\begin{comment} %--useful, verbose---
\paragraph{Handling control flow transfer}
While targets of all guest branch instructions are addresses of untranslated guest instructions,
the host instructions should transfer the control to corresponding host instructions in the code cache. 
To do so, the first time the DBT translates such a branch instruction, it emits a trampoline instruction transferring the control back to DBT. 
The DBT looks up the corresponding code cache address; 
if the target is untranslated yet, the DBT translates it now (so as the name ``dynamic translation''). 
Either way, the DBT will obtain the code cache address. 
The DBT patches the trampoline instruction with a direct jump to the code cache address. 
All subsequent execution of the TB will follow the direct jump to the code cache address without going through the DBT. 
\end{comment}

\paragraph{Design overview}
We build \sys{} atop QEMU~\cite{qemu}, a popular, opensource cross-ISA DBT engine. 
%one of the few open-source cross-ISA DBT
%we directly bridge the two ISAs (host and guest) so that the DBT engine:
\sys{} inherits QEMU's infrastructure but departs from its generic design which translates between arbitrary ISAs.
%translating an arbitrary guest ISA to an arbitrary host ISA. 
\sys{} targets two well-known DBT optimizations:
i) to emit as few host instructions as possible; 
%ii) to keep the execution at full speed in the code cache (without invoking the DBT engine) as much as possible.
%ii) to minimize invocations of the DBT engine.
ii) to exit from the code cache to the DBT engine as rarely as possible.
We exploit the following similarities between the CPU's and the \pmcore{}'s ISAs (ARMv7a \& ARMv7m):

\begin{myenumerate}
\item 
Most v7a instructions have v7m counterparts with identical or similar semantics, albeit in different encoding. (\S\ref{sec:isa:insn})
%(albeit in different binary formats). 
%Identical or similar instruction semantics. 
%A large portion of instructions share same or similar semantics (encodings or formats), allowing mapping one guest instruction to one or a few host instructions. 
\vspace{-1.5em}
\item 
%The number of general purpose register of host should be larger or at least equal to guest.
%Host (v7m) has the same number of general purpose registers as the guest (v7a). 
Both ISAs have the same general purpose registers.
The condition flags in both ISAs have same semantics. (\S\ref{sec:isa:reg})
%\item 
%The CPU flags in both ISAs have similar semantics and affect conditional instructions in similar ways. 
%Both ISAs have CPU flags that have similar semantics and affect conditional instructions in similar ways. 
%The condition flags in both ISAs have same semantics. 
%\item 
%Guest and host kernel should sit in the same address space or differ only by a constant offset. 
\vspace{-0.5em}
\item 
%Both ISAs use PC, SP, and LR in the same way.
%Both ISAs use PC, LR, and SP in the same way.
Both ISAs use program counter (PC), link register (LR), and stack pointer (SP) in the same way. (\S\ref{sec:isa:lrsp})
\end{myenumerate}

Beyond the similarities, the two ISAs have important discrepancies. 
Below, we describe our exploitation of the ISA similarities and our treatment for \textit{caveats}.

%Below, we describe our exploitation of the ISA similarities and 
%highlight the important \textit{caveats} that we have to fix.
%due to the ISA discrepancies. 

% !TeX root = main.tex

%\subsection{Automatic instruction mapping}
%\subsection{Exploiting Similar Instruction Encodings}
%\subsection{Exploiting Similar Instruction Semantics}
\subsection{Exploiting Similar Instruction Semantics}
\label{sec:isa:insn}

%XXX specifies the semantics of each instruction.

%Our approach to producing translation rules is more systematic and thorough than relying on developer's knowledge or learning from existing compiled executable binaries. \note{cite}

%Written in XML, it is organizes by instruction mnemonics (e.g. ADD); 
%under each mnemonic it lists all instruction encoding variants (e.g. ADD Rn.. and ADD ...), each of which belongs to one specific ISA (e.g. v7a or v8m). 
%Under the same mnemonic, instructions of the same ISA have different semantics (e.g. addition of two register values versus addition between ...); 
%yet, instruction in one ISAs often have counterparts in other ISA sharing same or similar semantics. 
%\note{showing a snippet? -- cut/compress}

% !TeX root = main.tex

% Please add the following required packages to your document preamble:
% \usepackage[table,xcdraw]{xcolor}
% If you use beamer only pass "xcolor=table" option, i.e. \documentclass[xcolor=table]{beamer}
%\begin{table}[]
\begin{wraptable}{L}{0.25\textwidth}
%\footnotesize
\centering
%\vspace*{-1em}
%	\begin{tabular}{ |c|c|c|}
%		\hline
%		 & number & description \\
%		 \hline
%		total insn & 497 & \\
%		\hline
%		A encoding & 558 & \\
%		\hline
%		T encoding & 617 & \\
%		\hline
%		direct mappable & 409 & same fields \\ 
%		\hline
%		direct mappable & 390 & same fields and field width \\
%		\hline
%		imm enc & 16 &  different imm encoding \\
%		\hline
%		imm width & 14 & imm8 vs imm12, etc\\
%		\hline
%		reg range & 10 & different reg bits \\
%		\hline
%		ctrl mode & 16 & ctrl bits difference \\
%		\hline			
%	\end{tabular}

% enc -- the imm encoding is different. have same # bits in imm
% width -- different encoding. and have diff # bits 
	
%	\begin{tabular}{ |c|c|c|c|}
%		\hline
%		 & Count & \# v7m & Example \\
%		\hline
%		Identical semantics & 447 & 1 & ADD\_R \\ 
%		\hline
%		Constraints on constant & 22 & 3-5 & STR\_i A1 vs T4 \\
%		\hline
%		Side effect & 52 & 2 -- 5 & \\ 
%		\hline
%		Richer shift modes & 10 & 2 & \\ 		
%		\hline
%		No counterpart & 27 & 2-5 & RSC\note{decide upper bound}\\
%		\hline
%		Total A & 558 & - & \\		
%		\hline
%	\end{tabular}	
	
	   	\includegraphics[width=0.25\textwidth]{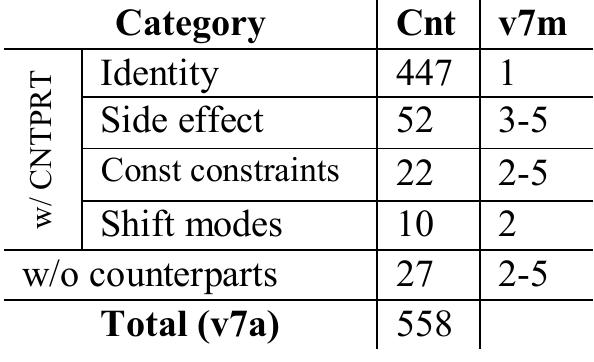}
	
\caption{Translation rules for v7a instructions. Column 3: the number of v7m instructions emitted for one v7a instruction}
\label{tab:insn}
\vspace*{-1em}
\end{wraptable}

We devise translation rules with a principled approach by 
parsing a machine-readable, formal ISA specification recently published by ARM~\cite{arm-spec}.
Our overall guideline is to map each v7a instruction to one v7m instruction that has identical or similar semantics. 
We call them \textit{counterpart} instructions. 
For a counterpart instruction with similar (yet non-identical) semantics, 
\sys{} emits a few ``amendment'' v7m instructions to make up for the semantic gap. 
The resultant translation rules are based on individual guest instructions, different from translation rules based on one or more translation blocks commonly seen in cross-ISA DBT~\cite{qemu-ml-rules}.
%This contrasts to DBT across ISA families (e.g. ARM and x86), which often require
%translation rules based on instruction blocks (basic blocks?) or even across instruction blocks (??)
%This is because ISA similarity allows DBT to translate most guest instructions to one host counterpart instruction, as we will show below.
This is because semantics similarities allows identity translation for most guest instructions.
Amendment instructions are oblivious to interrupts/exceptions:
as stated in \S\ref{sec:emu:irq}, ARK defers IRQ handling to translation block boundary and expects no exceptions. 
%We expect no exceptions. Details in .

Table~\ref{tab:insn} summarizes \sys{}'s translation rules for all 558 v7a instructions. 
Among them, 80\% can be translated with identity rules, 
for which \sys{} only needs to convert instruction encoding at run time.
%\st{Note that the binary format of the v7m instruction often differs from the v7a, which requires DBT to emit the v7m instruction at run time (which is true for one-to-one translation rules).} 
15\% of v7a instructions have v7m counterparts but may require amendment instructions, which % v7m instructions as amendment. 
fortunately fall into a few categories: % (summarized in Table~\ref{tab:insn}):
%For each v7a instruction, the analyzer searches under the same mnemonic for v7m instruction with same semantics. 
%If found, an identity translation rule between the two instructions is added. 
%Our analyzer identifies XXX of such rules (XX\% of all v7a instructions). 
%In the remainder of the v7a instructions, XXX (XX\%) have their v7m counterparts with similar semantics but need additional ``amendment'' instructions to be identical. 
%while the v7m instructions see extra constraints on instruction operands (often due to weaker decoding capabilities \note{fix}), 
%e.g. smaller immediate values, smaller constants for register shift, limited shift modes, general purpose registers from a smaller range (e.g. R0--R7).
%For them, the analyzer adds translation rules that check the immediate values or register indices (which are known at translation code time) and produce one-to-one (\textbf{identity}) translation if possible and otherwise produce 1 -- 5 instructions selected from a small library of code snippets
 % 1 for data processing insn with diff shift
 % 5 for conditional ld/st with large imm and wb?
%With the help from analyzer, we come up a systematic method to handle subtle difference in instructions between ARMv7 and ARMv7m.
%
%
i) \textbf{\textit{Side effects.}}
%Some v7a instructions has additional side effects on CPU state or memory compared to their v7m counterparts.
After load/store, v7a instructions
%compared to their v7m counterparts, 
may additionally update memory content or register values (shown in Table~\ref{tab:dbt-example}, G1). 
%Possible side effects include writing back an arithmetic instruction's result to memory (shown in Table~\ref{tab:dbt-example}\circled{2}) and writing back an updated index to a register. 
%and updating the index register used in the instruction.
%Possible side effects include writing back an arithmetic instruction's result to memory, 
%and updating the index register used in the instruction.
\sys{} emits amendment instructions to emulate the extra side effect (H3). 
ii) \textbf{\textit{Constraints on constants.}}
%The range of constant values that can be encoded in v7m instructions may see additional constraints on v7a. 
The range of constants that can be encoded in a v7m instruction is often narrower (Table~\ref{tab:dbt-example}, G2).
% -- verbose --- 
%the ADD instruction in v7a can encode constants in [0,255] while ADD in v7m can only encode constants in [128,255].
% -- useful ---
%the ADD instruction in v7a can encode constants in [0, 255] with right rotation by [0, 30] bits at increment step of 2, while ADD in v7m can only encode constants in [128, 255] with right rotation by [4, 31] bits.
%This also limits the memory range, which is expressed with a constant offset, that a v7m instruction can address. 
%For instructions followed in this category, \sys{} will check the range during translation.
%If the address in guest instruction is not addressable in single instruction host instruction,
In such cases, the amendment instructions load the constant to a scratch register, operate it, and emulate any side effects (e.g., index update) the guest instruction may have. 
iii) \textbf{\textit{Richer shift modes.}}
v7a instructions support richer shift modes and larger shift ranges than their v7m counterparts.
This is exemplified by Table~\ref{tab:dbt-example} G1, where  
a v7m instruction cannot perform LSR (logic shift right) inline as its v7a counterpart. 
%Similar to above, the amendment instructions load the operand to a scratch register and perform shift on the register. 
Similar to above, the amendment instructions perform shift on the operand in a scratch register.

% !TeX root = main.tex

% Please add the following required packages to your document preamble:
% \usepackage[table,xcdraw]{xcolor}
% If you use beamer only pass "xcolor=table" option, i.e. \documentclass[xcolor=table]{beamer}
\begin{table}[t]
%\footnotesize
\centering
%\vspace*{-1em}
   	\includegraphics[width=0.42\textwidth]{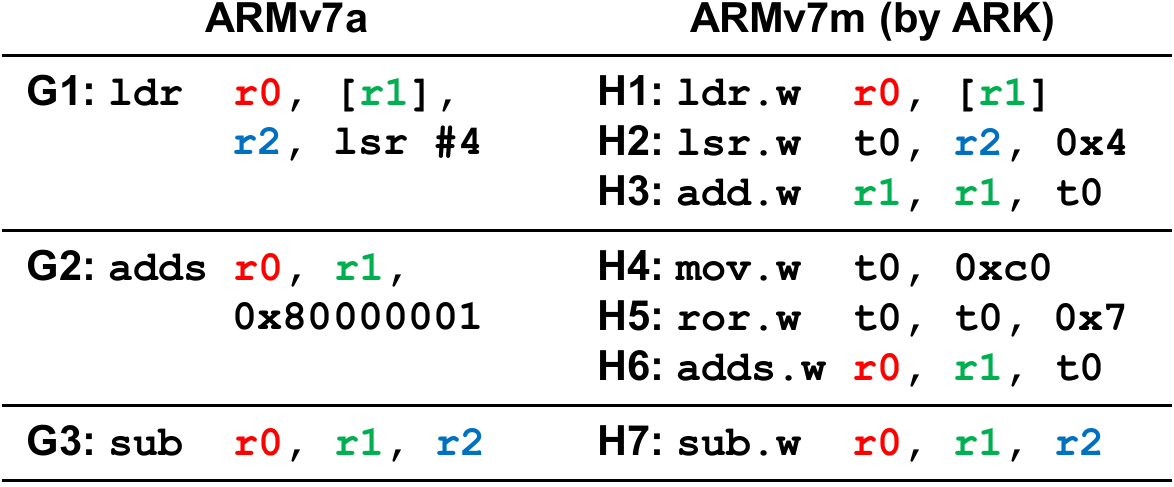}
	\caption{Sample translation by \sys{}. 
	By contrast, our baseline QEMU port translates G1--G3 to \textbf{27} v7m instructions}
	\label{tab:dbt-example}
\vspace*{-1.5em}
\end{table}

Beyond the above, only 27 v7a instructions have no v7m counterparts, for which we manually devise translation rules. 
%\note{cut?} The fraction of hand-written rules is much smaller than QEMU, in which all IR to host instructions rules are hand-written based on DBT author knowledge. 

In summary, through systematic exploitation of similar instruction semantics, \sys{} emits compact host code at run time. 
In the example shown in Table~\ref{tab:dbt-example}, three v7a instructions are translated into seven v7m instructions by \sys{}, while to 27 instructions by our QEMU baseline. % which results in much higher execution overhead.

\subsection{Passthrough of CPU registers}
\label{sec:isa:reg}

\paragraph{General purpose registers}
Both the guest (v7a) and the host (v7m) have 
the same set (13) of general-purpose registers. 
%In translating a guest instruction into its host counterpart, 
%\sys{} tries best to allocate the same host registers as in the guest instruction. 
% contrast to QEMU?
%By doing so, \sys{} reuses the register allocation done by the guest binary compiler.
%By doing so, \sys{} follows the register allocation decisions made by the compiler that generates the guest binary.
In allocating registers of a host instruction, 
\sys{} follows guest register allocation with best efforts (e.g., one-to-one mapping in best case, as in Table~\ref{tab:dbt-example}, G1).
%which is decided by the the compiler generating the guest binary.
%decisions made by the compiler for the guest binary.
\sys{} emits much fewer host instructions than QEMU, 
which emulates all guest registers in host memory with load /store. 
%To translate a guest instruction that operates on registers, 
%QEMU emits host instructions to load and store the corresponding variables in memory.  
%By doing so, \sys{} best reuses the register allocation done by the guest binary compiler; 
%while QEMU has to allocate host register at translation time based on its limited scope of TB. 
%\note{can we ref to table 3?}

%However, the register allocation method has important caveats.
%We fix important caveats in this register allocation method. 
\caveat{} 
The amendment host instructions operate scratch registers
%(called ``scratch register'' in DBT lingo),
as exemplified by \code{t0} in Table~\ref{tab:dbt-example}, H2-H6.
However, the wimpy host faces higher register pressure, as 
it (v7m) has no more registers than the brawny guest (v7a). 
%\sys{} has to spill some registers to memory. 
%the host has no spare ones as scratch registers. 
%This contrasts to many cross-ISA DBT where the host has richer registers than the guest~\cite{mambo-x64}.
%\sys{} has to spill some registers: swapping their contents to memory so that these registers can be temporarily used as scratch registers. 
To spill some registers to memory while still reusing the guest's register allocation,
we make the following tradeoff: 
we designate one host register as the \textit{dedicated} scratch register, and emulates its guest counterpart register in memory. % (as QEMU does to all the guest registers).  % which incurs expensive memory access. 
We pick the \textit{least} used one in the guest binary as the dedicated scratch register,
which is experimentally determined as R10 by analyzing kernel binary.
%we experimentally determined it as R10 by analyzing the ARM Linux kernel.
We find most amendment instructions are satisfied by \textit{one} scratch register; 
in rare cases when extra scratch registers are needed, 
\sys{} follows a common design to allocate dead registers and spill unused ones to memory.

\begin{comment} %---- useful but verbose ---- %
\note{the case we need temp regs}
1. Non-control related instruction which uses PC, and PC needs to be put to a temporary register because it points to guest address space and is only known in translation time.
%2. Width of immediate value is shorter on host, so immediate operand from guest instruction needs to be moved to a temporary register and covert to a register-typed operand.
%3. For 1:N translation, temporary registers may needed for additional instructions.
4. If host and guest are in different address space, an offset will be applied to memory address and the result will be stored in temporary register.
\end{comment}

%These extra host instructions need ``scratch'' registers.  \note{also point to the CPU struct}
%For this purpose, \sys{} preserves two least used registers (R10 and R11) in the host binary to be used as scratch registers. 
%Hence, the host's access to R10 and R11 is still emulated by software instead of the guest R10 and R11. 
%However, since few host instructions (\note{stats?}) access these two, \sys{} minimizes the overhead. 

\paragraph{Condition flags}
%\note{done cutting}
Both the guest and the host ISAs involve five hardware condition flags (e.g., zero and carry) with identical semantics; 
fortunately, most guest (v7a) instructions and their host (v7m) counterparts have identical behaviors in testing/setting flags per the ISA specifications~\cite{arm-spec}. %most guest (v7a) instructions and their host (v7m) counterparts have identical behaviors in testing/setting flags per the ISA specifications~\cite{arm-spec}. %
\sys{} hence directly emits instructions to manipulate the host's corresponding flags.
%Both the guest and the host ISAs involve five hardware condition flags (e.g. zero and carry) with identical semantics.
%\sys{} hence directly emits instructions to manipulate the host's corresponding flags.
%Fortunately, most guest (v7a) instructions and their host (v7m) counterparts have identical behaviors in testing/setting flags per the ISA specifications~\cite{arm-spec}. %
%influence/dependency on condition flags, 
%verified by us parsing the ISA specifications~\cite{arm-spec}. % (\S\ref{sec:isa:insn}).
%affect and depend on the flags in the same way
%Such flag passthrough hence incurs much lower overhead than QEMU, which emulates guest CPU flags as individual variables in host memory and emits as many as 7 host instructions to manipulate emulated flags, even for each flag-setting guest instruction.
Such flag passthrough especially benefits control-heavy \snr{}, which contains extensive conditional branches (\S\ref{sec:bkgnd});
we study its benefits quantitatively in \S\ref{sec:eval:exec}.
%hence incurs much lower overhead than QEMU, which emulates guest CPU flags as individual variables in host memory and emits as many as 7 host instructions to manipulate emulated flags, even for each flag-setting guest instruction.
%For each conditional guest instruction, QEMU emits host instructions for loading the tested flags. 
%This cost is especially high for the control-heavy \snr{}, which contains extensive conditional branches (\S\ref{sec:bkgnd}). 

\caveat{} 
%The caveat, again, lies in amendment host instructions, which may affect the host hardware (native?) CPU flags unexpectedly. 
Amendment host instructions may affect the hardware condition flags unexpectedly. 
%Hence, 
%\sys{} disables an amendment instruction's influence on condition flags when possible; 
For amendment instructions (notably comparison and testing) that \textit{must} update the flags as mandated by ISA, \sys{} emits two host instructions to save/restore the flags in a scratch register around the execution of these amendment instructions. 

\subsection{Control Transfer and Stack Manipulation}
\label{sec:isa:lrsp}

%\paragraph{PC and LR}
\paragraph{Function call/return}
%Both v7a and v7m use the two CPU registers for control flow: PC for program counter and LR for linked register (saving the function return address). 
%Program counter (PC) and function return address (LR) are two dedicated ARM registers crucial to control flow.
%In both v7a and v7m, PC serves as program counter and LR saves return address. 
%Both guest (v7a) and host (v7m) use PC (program counter)
Both guest (v7a) and host (v7m) use PC (program counter) 
%and LR (link register, storing function return address) 
and LR (link register) 
to maintain the control flow. 
QEMU emulates guest PC and LR in host memory.
%For each kernel function call and return,
%the emitted ARK code manipulates the emulated PC and LR.
%it pushes the emulated PC/LR values to stack and loads the emulated PC to emulated LR. 
%This ensures transparency: (host) code cache addresses is never leaked to the emulated PC, LR, or stack.
%However, this is expensive since the return address, loaded from stack or the emulated LR, points to a guest address.
As a result, the return address, loaded from stack or the emulated LR, points to a guest address (i.e., kernel address).
Each function return hence causes the DBT to step in and look up the corresponding code cache address. 
%Each return instruction is therefore translated into XXX host instructions. (??)
This overhead is magnified in the control-heavy \iophase{}.

%Hence, \sys{} only maintains an emulated PC in order to discover untranslated guest instructions. 
%Hence, \sys{} only track the guest PC in order to discover untranslated guest instructions. 
By contrast, \sys{} never emits host code to emulate the guest (i.e., kernel) PC or LR. 
For each kernel function call, ARK saves the return addresses within \textit{code cache} on stack or in LR; 
for each kernel function return,
ARK loads the return address (which points to code cache) to hardware PC from the stack or the hardware LR. 
%By doing so, guest function return does not invoke the DBT engine.
By doing so, \sys{} no longer participates in all function returns. 
Our optimization is inspired by same-ISA DBT~\cite{kernel-dbt}. 

\paragraph{Stack and SP}
QEMU emulates the guest (i.e., kernel) stack and SP with a host array and a variable. %in with an array and a program variable as the stack pointer (SP). 
%For instance, for each guest push, it updates the emulated SP variable and saves values to the array. 
Each guest push/pop translates to multiple host instructions updating the stack array and the emulated SP.
This is costly, as \snr{} frequently makes function calls and operates stack heavily.

\sys{} avoids such expensive stack emulation by emitting host push/pop instructions to directly operate the guest stack \textit{in place}. 
This is possible because \sys{} emulates the Linux kernel's virtual address space (\S\ref{sec:emu:workflow}).
\sys{} also ensures the host code generate the same stack frames as the guest would do by 
making amendment instructions avoid using stack, which would introduce extra stack contents.
In addition, this further facilitates the migration in abort (\S\ref{sec:abort}). 

\caveat{}
i)
As the host saves on the guest stack the code cache addresses, which are meaningless to the guest CPU,   
upon migrating from the \pmcore{} (host) to the CPU (guest), the DBT rewrites all code cache addresses on stack with their corresponding guest addresses.
ii) 
guest push/pop instruction may involve emulated registers (i.e., scratch register). 
%If one of these guest registers (e.g. R10) coincides with the host scratch register, 
%If one of these guest registers is emulated (i.e., its host counterpart is a scratch register), 
\sys{} must emit multiple host instructions to correctly synchronize the emulated registers in memory. 
\section{Translated $\longrightarrow$ Native Fallback}
\label{sec:abort}

As described in \sect{overview}, when going off the hot paths, \sys{} migrates the kernel phase back to the CPU and continues as native execution, 
analogous to virtual-to-physical migration of VMs~\cite{vmware-v2p}.
Migrating one DBT context is natural, as \sys{} passes through most CPU registers and uses the kernel stack in place (\S\ref{sec:isa:lrsp}). 
%the translated execution uses the kernel (guest) stack in place and \sys{} only has to rewrite all the code cache addresses on the stack as described in \sect{isa:lrsp}. 
Yet, to migrate \textit{all} active DBT contexts, \sys{} address the following unique challenges. 

%transfers the control (all DBT contexts) to the kernel which will finish the remaining \snr{} transparently.
%%The objective is all the unfinished kernel path and deferred functions can 
%We call this procedure XXXX. 

\begin{comment}
%Transfer the architectural state of one dispatcher (DBT) context to the CPU?
\paragraph{Migrate the architectural state}
\sys{} copies the host registers of the \pmcore{} to the CPU. 
%Because DBT translates all function calls, stack frame is preserved and can be directly used for big core. \note{fix}
Since the \sys{} DBT directly uses guest stacks in place, it passes back the SP, and rewrites all the code cache addresses saved on the stack as described in \sect{isa:lrsp}. 
\note{still valid?}
To solve these issues, \sys{} iterates through (unwind) the stack frames; 
consulting the translation information kept by the dispatcher, it rewrites the host code addresses with the corresponding guest addresses with its DBT knowledge. 
%This is done by mapping each host instruction back to TB, and then finding the corresponding guest instruction address.
\end{comment}

\paragraph{Migrate DBT contexts for deferred work}
%how to migrate DBT contexts for executing deferred functions?
%While migrating the main (master?) context is natural \note{explain?}. 
%Upon migration, these contexts may be blocked in deferred work, which should continue to execute on the CPU.
After fallback, all blocked workitems should continue their execution on the CPU. 
Unfortunately, their enclosing DBT contexts do not have counterparts in the Linux kernel. 
To solve this issue, we again exploit the insight that the workitems are oblivious to their execution contexts. 
Upon migration, the Linux kernel creates temporary kernel threads as ``receivers'' for blocked workitems to execute in. 
%The kernel threads participate process scheduling by the kernel. 
Once the migrated workitems complete, the receiver threads terminate.
%Any new deferred functions are scheduled with the kernel's own mechanism. 

%Third, 
%how to migrate an interrupt context. 
\paragraph{Migrate DBT context for interrupt}
If fallback happens inside an ISA-neutral interrupt handler (translated), the remainder of the handler should migrate to the CPU.
This challenge, again, is that \sys{}'s interrupt context has no counterpart on the CPU: the interrupt never occurs to the CPU. 
\sys{} addresses this by \textit{rethrowing} the interrupt as an IPI 
(inter-processor interrupt) from the \pmcore{} to the CPU; 
the Linux kernel uses the IPI context as the receiver for the migrated interrupt handler to finish execution. 
%executing from the fallback point in the interrupt handler.
%the IPI handler (in its interrupt context) of the CPU executes the early, ISA-specific interrupt handling (as described in \sect{sys}) was the entrance to kernel's ISA-neutral interrupt handling and EKC's DBT started. 
%Then, the kernel loads the emulated GP regs. 

% --- xzl: this hurts us --- XXX: add it back later %
%\sys{} gives up complete transparency: if Linux checks the CPU's interrupt controller, it will find that the interrupt comes from IPI instead of an IO device. 

% -- verbose -- 
%if Linux checks its hardware interrupt controller, it will be confused by observing that the interrupt comes from IPI instead of an IO device. 
%However, we do not observe such a behavior, as the interrupt handling has already finished its early stage in which it has read the interrupt type. 

Section~\ref{sec:eval} will evaluate the fallback frequency and cost.

%There are two problems introduced by our optimizations:
%1). Since \sys{} directly uses address of code cache for control flow, the return address for each frame on stack points to translated code.
%2). 
%We proposed solutions as following:
%1). Before migrating to powerful core, \sys{} will iterate through stack and replace all return addresses.
%Since function calls define the boundary of TBs, \sys{} can determine return address by adding 4(because all ARM's instructions are 4 bytes aligned) to last instruction's address.
%2). Since stack is contiguous in virtual memory, \sys{} will add an offset for all pointers pointing to stack. 

% --- +2=8 page --- %
%\input{impl}
% !TeX root = main.tex

\section{Evaluation}
\label{sec:eval}

We seek to answer the following questions:
\begin{myenumerate}
%\item Does \sysmodel{} reduce the current and future engineering effort? 
\item Does \sys{} incur tractable engineering efforts?  (\S\ref{sec:eval:code})
\vspace{-0.5em}
%\item Does \sys{} incur low execution overhead? (\S\ref{sec:eval:exec})
\item Is \sys{} correct and low-overhead? (\S\ref{sec:eval:exec})
%\item Does \sys{} reduce energy for \snr{}? How would this impact platform battery life? 
\vspace{-0.5em}
\item Does \sys{} yield energy efficiency benefit? 
What are the major factors impacting the benefit? (\S\ref{sec:eval:energy})
%\item What are the major factors contributing to the energy benefit? 
%\item How much does the overhead matter to energy benefit?
%\item How much overhead does \sys{} incur in suspend/resume translation?
%\item How does ISA similarity affect translation overhead?
%\item Where does the energy benefit come from? 
%\item What are the major factors impacting the energy benefit? 
\end{myenumerate}

\subsection{Methodology}

\paragraph{Test Platform}
%We evaluate \sys{} on Pandaboard-ES Rev B1 equipped with TI OMAP4460 SoC~\cite{omap4rtm} \note{just mention it is an A9 board}.
We evaluate \sys{} on OMAP4460, an ARM-based SoC~\cite{omap4rtm} as summarized in Table~\ref{tab:hw}. 
%it has dual ARM Cortex-A9 CPU and dual Cortex-M3 as summarized in Table~\ref{tab:hw}. 
%We run \sys{} on one Cortex-M3. 
%Among a variety of platforms that fit our hardware model (\S\ref{sec:bkgnd:hw}), we chose this SoC for its good documentation and long-time kernel support (since 2.6.11), which allows our study of kernel ABI over a long timespan in \sect{bkgnd}. % (in Table~\ref{tab:abi}). 
We chose this SoC mainly for its good documentation and longtime kernel support (since 2.6.11), which allows our study of kernel ABI over a long timespan in \sect{bkgnd}. % (in Table~\ref{tab:abi}). 
As Cortex-M3 on the platform is incapable of DVFS, for fair comparison, we run both cores at their highest clock rates. 
Note that OMAP4460 is not completely aligned with our hardware model, for which we apply workarounds as will be discussed in \sect{eval:discussion}.

%\input{tab-devboards}

%\paragraph{Benchmark}
%\paragraph{Test setup}
\paragraph{Benchmark setup}
%We evaluate \sys{} and report results from two aspects: 1) translation overhead, 2) emulation costs. 
We benchmark \sys{} on the whole suspend/resume kernel phases. 
We run a user program as the test harness that periodically kicks ARK for suspend/resume; 
the generated kernel workloads are the same as in all ephemeral tasks.
Our benchmark is macro: it exercise extensive drivers and services, during which ARK translates and executes over 200 million instructions. 

%We test \sys{} with Linux kernel v4.4. 
%We configure the platform so that the kernel operates nine devices for \snr{}.
%their \snr{} involves complex kernel code; 
%together, their exercise extensive kernel services. 
The benchmark operates nine devices for \snr{}.
\begin{enumerate*}
	\item \textbf{SD card}: SanDisk Ultra 16GB SDHC1 Class 10 card;
	\item \textbf{Flash drive}: a generic drive connected via USB;
	\item \textbf{MMC controller}: on-chip OMAP HSMMC host controller; % involving register read/write;
	\item \textbf{USB controller}: on-chip OMAP HS multiport USB host controller; % involving deferred work;
	\item \textbf{Regulator}: TWL6030 power management IC connected via I2C; % manage power for hardware components;  % involving I2C bus communication;
	\item \textbf{Keyboard}: Dell KB212-B keyboard connected via USB;  % involving deferred work;
	\item \textbf{Camera}: Logitech c270 connected via USB; % involving deferred work;
	\item \textbf{Bluetooth NIC}: an adapter with Broadcom BCM20702 chipset connected via USB;
	\item \textbf{WiFi NIC}: TI WL1251 module. 
%	involving page allocation, DMA, SDIO, and softirq handling; 
\end{enumerate*}
The kernel invokes sophisticated drivers, thoroughly exercising various services including
deferred work (2--4,6--8), slab/buddy allocator (1--4,6--9), softirq (9), DMA (2,6--9), threaded IRQ (1,5,9), and firmware upload (9).

% --- old ones --- 
%\begin{enumerate}
%	\item MMC sd: MMC command sending/handling.
%	\item MMC host controller: host read/write.
%	\item USB Keyboard: generic USB framework
%	\item USB Flash Drive: handling USB storage devices.
%	\item USB Camera: \todo{todo}.
%	\item USB host: \todo{todo}
%	\item Bluetooth: among the most frequently exercised driver..
%	\item Regulator: I2C bus communication
%	\item WiFi: sdio 
%\end{enumerate}

%(1). MMC: This is a common storage media on mobile/IoT devices, including SD card or eMMC.
%During suspend, it will send commands to put card into idle status, and then adjust supply voltage for MMC device to low power state.
%(2). USB Keyboard: This benchmark exercise translation overhead for generic USB devices.
%It will drain all pending URBs(USB Request Block) and unregister the device during suspend.
%In resume, driver will register the device, send command to check port status.
%This will also verify interrupt handling mechanism.
%(3). USB Flash Drive:
%This benchmark uses the same USB suspend/resume framework.
%However, it has additional hooks to serve USB storage devices.
%(4). Bluetooth:
%(5). Regulator:

We measure device \snr{} executed by \sys{} on Cortex-M3 
and report the measured results.
We compare \sys{} to native Linux execution on Cortex-A9. 
We further compare to a baseline \sys{} version: its DBT is a straightforward v7m port of QEMU that misses 
optimizations described in \sect{dbt}.
We report measurements taken with warm DBT code cache, 
as this reflects the real-world scenario where device \snr{} is frequently exercised.

%we execute \sys{} once to warm up its DBT code cache, and report the measurement from subsequent executions. 
%Based on our assumption, suspend/resume should be frequently executed, therefore the initial translation overhead could be amortized by subsequent execution.

\subsection{Analysis of engineering efforts}
\label{sec:eval:code}
% !TeX root = main.tex

% Please add the following required packages to your document preamble:
% \usepackage[table,xcdraw]{xcolor}
% If you use beamer only pass "xcolor=table" option, i.e. \documentclass[xcolor=table]{beamer}
%\begin{wraptable}[]
\begin{wraptable}{R}{0.2\textwidth}
%\footnotesize
\centering
\vspace*{-1em}
   	\includegraphics[width=0.2\textwidth]{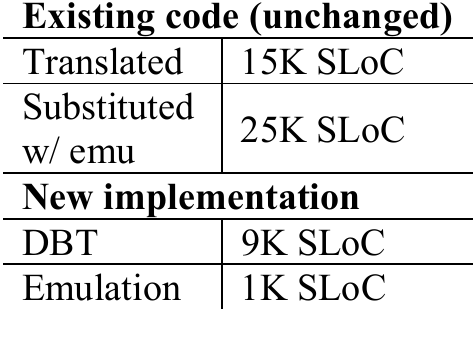}
  	\vspace{-20pt}
   	\caption{Source code}
   	\label{tab:eval-sloc}
\vspace{-10pt}
\end{wraptable}

% BINARY: 1036kb IN total

% DBT 8840
% Emulated Kernel core 710
% Misc 350 (script)
% Trasnlated code: 15k

\sys{} eliminates source refactoring of the Linux kernel (\S\ref{sec:bkgnd:sw}).
As shown in Table~\ref{tab:eval-sloc}, \sys{} transparently reuses substantial kernel code (15K SLoC in our test), most of which are drivers and their libraries.
%\sys{} hence eliminates the engineering effort in porting and rebuilding them for the \pmcore{}. 
We stress that \sys{}, as a driver-agnostic effort, not only enables reuse of the drivers under test but also other drivers in the ARMv7 Linux kernel. 

Table~\ref{tab:eval-sloc} also shows that 
\sys{} requires modest efforts in developing new software for the \pmcore{}. 
The 9K new SLoC for DBT is low as compared to commodity DBT (e.g., QEMU has 2M SLoC). 
\sys{} implements emulation in as low as 1K SLoC and in return avoids translating generic, sophisticated Linux kernel services~\cite{kernel-dbt,drk}.
The result validates our principle of specializing these emulated services.

\sys{} meets our goal of ``build once, run with many''.
We verify that the \sys{} binary works with a variety of kernel configuration variants (including \code{defconfig-omap4} and \code{yes-to-all}) of Linux 4.4.
We also verify that \sys{} works with a wide range of Linux versions, from version 3.16 (2014) to 4.20 (most recent at the time of writing). % \note{check 4.20?}
This is because \sys{} only depends on a narrow ABI shown in Table~\ref{tab:ker-api}, which has not changed since Linux 3.16.

\subsection{Measured execution characteristics}
\label{sec:eval:exec}
% !TeX root = main.tex

\begin{figure}[t]
    \centering
   	\includegraphics[width=0.45\textwidth]{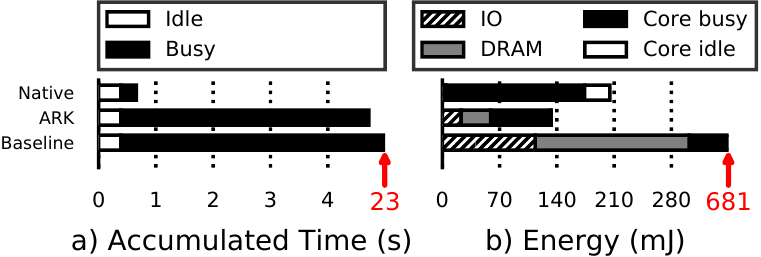}
%        \vspace*{-15pt}
    \caption{Execution time and energy in device \snr{}. \sys{} substantially reduces the energy.}
    \label{fig:eval-savings}
%    \vspace*{-15pt}
\end{figure}

%We profile the \sys{} execution. 
% read and write per seconds, 

\paragraph{ARK's correctness}
Formally, we derive translation rules from the specification of ARM ISA~\cite{arm-spec};
experimentally, we validate ARK by comparing its execution results side-by-side with native execution and examining the translated code with the native kernel binary.
Over 200 million executed instructions, we verify that \sys{}'s translation preserves kernel's semantics and presents consistent execution results. 

%\note{correctness -- as introduced before, besides devising translation rules from the book, we validate ARK by comparing its benchmark results side-by-side with native execution...}

\paragraph{Core activity}
We trace core states during ARK execution.
Figure~\ref{fig:eval-savings} (a) shows the breakdown of execution time. 
Compared to the native execution on CPU, \sys{} shows the same amount of accumulated idle time but much longer (16$\times$) busy time. 
The reasons are Cortex-M3's much lower clock rate ($1/6$ of the A9's clock rate) and \sys{}'s execution overhead. 
%The reasons are that one cycle on Cortex-M3 's cycle is is only 17\% of the CPU and \sys{}'s DBT overhead. 
Despite the extended busy time, \sys{} still yields energy benefit, as we will show below.
%the baseline exhibits even higher busy execution due to higher DBT overhead. 

% !TeX root = main.tex

\begin{figure}[t]
    \centering
   	\includegraphics[width=0.45\textwidth]{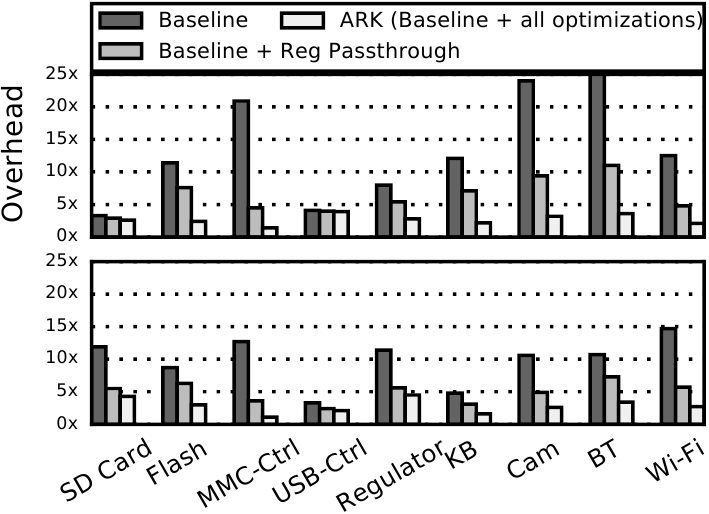}
    \caption{Busy execution overhead for devices under test (top: suspend; bottom: resume). Our DBT optimizations reduce the overhead by up to one order of magnitude}
    \label{fig:eval-dbt}
%    \vspace*{-15pt}
\end{figure}

\paragraph{Memory activity}
We collect DRAM activities by sampling the hardware counters of the SoC's DDR controller. 
We observed that \sys{} on Cortex-M3 generates much higher average DRAM utilization (32 MB/s read and 2MB/s write) than the native execution on A9 (only 8MB/s read and 4MB/s write). 
%The main reason is the cache asymmetry between the CPU and the \pmcore{}. 
We attribute such thrashing to M3's small (32KB) last-level cache (LLC).
Throughout the test, the \sys{} emitted and executed around 230KB host instructions, which far exceeds the LLC capacity and likely causes thrashing. 
%The touched kernel data is small, less than \note{tens of} KB. 
By contrast, Cortex-A9 has a much larger LLC (1MB), which absorbs most of the kernel memory access. 
The memory activity has a strong energy impact, as will be shown below.

%\subsection{ISA Similarities}
%\subsection{Validation of DBT designs}
%\subsection{DBT}

%\paragraph{DBT Overhead}
%\paragraph{Execution breakdown}
\paragraph{Busy execution overhead}
%\label{sec:eval-features}
Our measurement shows that \sys{} incurs low overhead in busy kernel execution, which includes both DBT and emulation. 
We report the overhead as the ratio between \sys{}'s cycle count on Cortex-M3 to the Linux's cycle count on A9. 
%Note that the cycle lengths in time differ by XXX are different clock rates. 
Note that an M3 cycle is 6$\times$ longer than A9 due to different clock rates. 

Overall, the execution overhead is 2.7$\times$ on average (suspend: 2.9$\times$; resume: 2.6$\times$). 
Of individual drivers, the execution overhead ranges from 1.1$\times$ to 4.5$\times$ as shown in Figure~\ref{fig:eval-dbt}.
%We notice the drivers with denser control transfer (e.g. USB) experience higher overhead due to higher DBT cost.
%
Our DBT optimizations (\S\ref{sec:dbt}) have strong impact on lowering the overhead.
Lacking them, our baseline design incurs a 13.9$\times$ overhead on average, 5.2$\times$ higher than \sys{}. 
We examined how our optimizations contribute to the gap: register passthrough (\S\ref{sec:isa:reg}) reduces the baseline's overhead by 2.5$\times$ to 5.5$\times$. 
Remaining optimizations (e.g., control transfer) collectively reduce the overhead by additional 2$\times$. 
Our optimizations are less effective on drivers with very dense control transfer (e.g., USB) due to high DBT cost.

\paragraph{Emulated services}
Our profiling shows that \sys{}'s emulated services incur low overhead. 
Overall, the emulated services only contribute 1\% of total busy execution.
i) The early, core-specific interrupt handling (\S\ref{sec:emu:irq}) takes 3.9K Cortex-M3 cycles, only 1.5--2$\times$ more cycles than the native execution on A9. 
ii) Emulated workqueues (\S\ref{sec:emu:tasklet}) incurs a typical queueing delay of tens of thousands M3 cycles. The delay is longer than the native execution but does not break the deferred execution semantics. 
%iii) For migrating one DBT context to the CPU in fallback, \sys{} spends around 20 us on rewriting code cache addresses on stack (\S\ref{sec:isa:lrsp}), 17 us to flush Cortex-M3's cache, and 2 us to wake up the CPU through an IPI.

\paragraph{Fallback frequency \& cost} 
We stress test \sys{} by repeating the benchmark for 1000 runs.
Throughout the 1000 runs, \sys{} encounters only four cases when the execution goes off the hot path, 
all of which caused by the WiFi hardware failing to respond to resume commands; 
it is likely due to an existing glitch in WiFi firmware. 
%the Wi-Fi resume process and are attributed to a command sending failure.
In such a case, \sys{} migrates execution by spending around 20 us on rewriting code cache addresses on stack (\S\ref{sec:isa:lrsp}), 17 us to flush Cortex-M3's cache, and 2 us to wake up the CPU through an IPI.

\begin{comment} % --- useful --- -
\begin{enumerateinline}
\item 
\textbf{IRQ handling}
Compare with main CPU with hardware IRQ handling, \sys{} emulation approach takes 1.5 to 2$\times$ cycles from detecting IRQ to enter interrupt handler.
The latency could be higher depending on whether context switch is allowed.
\item 
\textbf{Workqueue} 
We measure the latency of from submission to finishing single work item, which can take tens of thousands cycles.
Although this could be longer than executing on CPU, we do not break the semantics of deferred execution.
\item 
\textbf{Fallback}
We run a micro benchmark and test the fallback mechanism.
It takes 20us to patch return addresses with 4 stack frames, 17us to flush cache and 2us to send out IPI.
 
\end{enumerateinline}
\end{comment}

%\subsection{Results for \snr{}}
%\subsection{End-to-end results}
%\subsection{Energy efficiency benefits}
%\subsection{Modeled Energy benefits}
\subsection{Energy benefits}
\label{sec:eval:energy}

% !TeX root = main.tex

\begin{table}[t]
\footnotesize
%\scriptsize
\centering
\begin{comment}

\begin{tabular}{lll}
\hline
 & Busy & Idle \\ \hline
 CPU ~\cite{am43xxpet}  & 630mW @ 1.2GHz &  80mW @ 600MHz \\ \hline
 \Pmcore{}(pcore) ~\cite{am572xpet} & 17mW @ 200MHz & 1mW @ 200MHz \\ \hline
 \multirow{2}{*}{DRAM \cite{lpddr2}}  & 6mW for CPU& 1.3mW Self Refresh \\
 & 8.4mW pcore & \\
\hline
\end{tabular}
\end{comment}

% --- old table, has power numbers ---- %
\begin{comment}
\begin{tabularx}{0.48\textwidth}{XX}
\hline
 CPU & \Pmcore{}  \\ \hline
 Cortex A9 @ 1.2GHz & Cortex M3 @ 200MHz\\
 L1:64KB + L2:1MB  & L1:32KB \\
% $P_{cpu\_busy/idle}$:630mW/80mW ~\cite{am43xxpet}& $P_{pc\_busy/idle}$:17mW/1mW  ~\cite{am572xpet} \\ 
 $P_{cpu\_busy/idle}$:630mW/80mW & $P_{pc\_busy/idle}$:17mW/1mW \\ 
 % --- useful --- 
% \multirow{2}{*}{DRAM} & 8MB/s read, 4MB/s write & 32MB/s read, 2MB/s write \\
% & Power:3.8mW/1.3mW~\cite{lpddr2} & Power:8.4mW/1.3mW~\cite{lpddr2} \\
\hline   
\end{tabularx}
%Power numbers: busy/idle.
\begin{flushleft}
DRAM power model: Micron LPDDR2~\cite{lpddr2}; self-refresh ($P_{mem\_sr}$): 1.3mW.
IO average power ($P_{io}$): 5mW~\cite{central-pm}
\end{flushleft}
%\caption{Power model parameters}
\vspace{-1em}
\caption{The test platform}
\label{tab:power-params}
\label{tab:hw}
\vspace{-1em}
\end{table}
\end{comment}

% --- new table ---
%\footnotesize

\begin{tabularx}{0.48\textwidth}{lll}
\hline
&  CPU & \Pmcore{}  \\ \hline
Core & Cortex A9@1.2GHz & Cortex M3@200MHz\\
Cache & L1:64KB + L2:1MB  & L1:32KB\\
Typical busy/idle power& 630mW/80mW &17mW/1mW\\  
\hline   
\end{tabularx}
%Power numbers: busy/idle.
%\vspace{-1em}
\caption{The test platform - OMAP4460 on a Pandaboard}
\label{tab:power-params}
\label{tab:hw}
\vspace{-2em}
\end{table}

\paragraph{Methodology}
%\sout{We model the system power based on measured hardware activities.
%We choose modeling because i) our test platform is a development board with unoptimized board-level power; 
%ii) the board does not provide measurement points for DRAM power~\cite{measure-power}.}
%, which has strong impact as suggested by the measurement above. \note{improve}
%i) was not fabricated with the state-of-the-art process and may not reflect the efficiency; 
%\st{i) was fabricated with 45nm process, which does not reflect the efficiency of current process (10 nm);}
We study system-level energy and in particular how it is affected by \sys{}'s its extended execution time.
We include energy of both cores, DRAM, and IO.

%three major components that \sys{} has direct impact on. 
%two major components contributing dominant energy of the embedded platforms we target during \snr{}. \note{cite}
%The other power, e.g. IO devices, is much lower-power (XXX) on the embedded SoC.
% --- verbose --- 
%As Cortex-M3 on our test platform is incapable of DVFS, for fair comparison, we assume both cores run at their highest clock rate; 
%however, if M3 is capable of DVFS, the achieved energy benefits are even higher.

We measure power of cores by sampling the inductors on the power rails for the CPU and the peripheral core.
%we model core power as a function of core activities with TI's modeling tool~\cite{am43xxpet,am572xpet};
As the board lacks measurement points for DRAM power~\cite{measure-power},
we model DRAM power as a function of DRAM power state and read/write activities, with Micron's official power model for LPDDR2~\cite{lpddr2}.
%These power tools are official and have production quality.
The system energy of \sys{} is given by:

\begin{footnotesize}
$ E_{\sys{}} = \underbrace{{E_{core}}}_\text{Measured} + T_{idle} \cdot \underbrace{(P_{mem\_sr} + P_{io})}_{Modeled} + T_{busy} \cdot \underbrace{( P_{mem} + P_{io})}_{Modeled} $
\end{footnotesize}

%\begin{footnotesize}
%	$ E_{\sys{}} = \underbrace{{E_{core}}}_\text{Measured} + T_{idle} \cdot \underbrace{(P_{mem\_sr} + P_{io})}_{Modeled} + T_{busy} \cdot F \cdot C \cdot \underbrace{( P_{mem} + P_{io})}_{Modeled} $
%\end{footnotesize}

Here, $E_{core}$ is the measured core energy. 
All $T$s are measured execution time.
%All $T$s are elapsed time measured in native execution.
%$ F $ captures the ratio between the frequencies of CPU and the \pmcore{}.
%$ C $ is the average measured overhead in busy execution reported earlier, as the ratio between cycle counts of \sys{} and the native execution.
$P$s are power consumptions for DRAM and IO:
$P_{mem}$ is DRAM's active power derived from measured DRAM activities as described in \sect{eval:exec};
$P_{mem\_sr}$ is DRAM's self-refresh power, 1.3mW according to the Micron model;
$P_{io}$ is the average IO power which we estimate as 5mW based on prior work~\cite{central-pm}. 
Note that during \snr{}, IO devices no longer actively perform work, thus consuming much less power.

\paragraph{Energy saving}
\sys{} consumes 66\% energy (a reduction of 34\%) of the native execution, despite its longer execution time. 
The energy breakdown in Figure \ref{fig:eval-savings}(b)
shows the benefit comes from two portions:
i) in busy execution, \sys{}'s energy efficiency is 23\% higher than the native execution due to low overhead (on average 2.7$\times$); 
%(note that this is \textit{after} the DBT overhead). 
ii) during system idle, \sys{} reduces system energy to a negligible portion, as the \pmcore{}'s idle power is only 1.25\% of the CPU's.
Figure~\ref{fig:eval-savings}(b) highlights the significance of our DBT optimizations: 
the baseline, like \sys{}, benefits from lower idle power as well; however its high execution overhead ultimately leads to 5.1$\times$ energy compared to the native execution. 
%\textbf{Recommendation to architects:}
%\textbf{Implications on hardware design:}
Interestingly, \sys{} consumes \textit{more} DRAM energy than the native execution.
We deem the cause as Cortex-M3's tiny LLC (32KB) as describe earlier.
Our result suggests that the current size is suboptimal for the offloaded kernel execution. 
%We will discuss the implications on hardware in \sect{eval:discussion}. 

% !TeX root = main.tex

\begin{figure}[t]
    \centering
   	\includegraphics[width=0.45\textwidth]{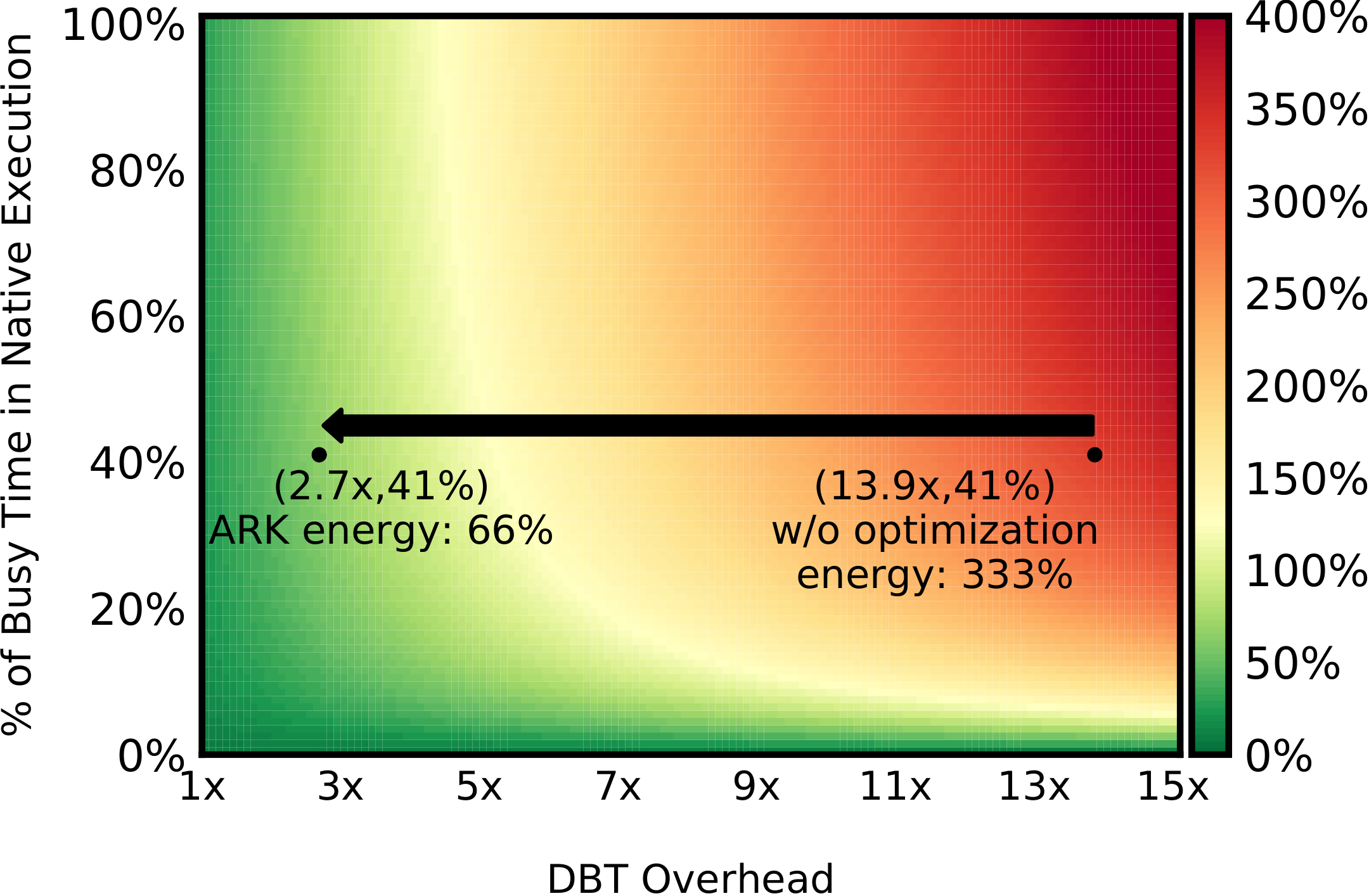}
    \vspace{-5pt}
    \caption{System energy consumption (inc. cores, DRAM, and IO) of \sys{} relative to native execution (100\%), under different DBT overheads (x-axis) and processor core usage (y-axis). 
    \sys{}'s low energy hinges on low DBT overhead.    
   % \textbf{XXX: }
    }
    \label{fig:heatmap}
    \vspace{-15pt}
\end{figure}

\paragraph{What-if analysis}
%We further show how the \sys{} energy benefit depends two major factors: the DBT overhead (\sys{}'s  behavior) and the fraction of busy execution time (Linux's behavior). 
%We further show the sensitivity of \sys{}'s energy benefit to two major factors: the DBT overhead (\sys{}'s  behavior) and the fraction of busy execution time (Linux's behavior). 
How sensitive is \sys{}'s energy saving to two major factors: the DBT overhead (\sys{}'s  behavior) and the processor core usage (Linux's behavior)?
To answer the question,
we estimate the \textit{what-if} energy consumption by using the power model as described above. 
%and plugging in different values of the two factors. 
% --- this is bad but we don't know how to fix --- 
% \note{fix?} We keep the DRAM activities constant, same as our measured values. 
The analysis results in Figure~\ref{fig:heatmap} show two findings. 
i) \sys{}'s energy benefit will be more pronounced with lower core usage (i.e., longer core idle), 
because \sys{}'s efficiency advantage over native execution is higher during core idle.
%and the DBT overhead is lower. 
%This is because a \pmcore{} excels at idle efficiency; 
%the efficiency advantage of a \pmcore{} over the CPU is higher during idle periods than during busy execution periods. 
%
%The analysis also highlights 
ii) \sys{}'s energy benefit critically depends on DBT. 
When the DBT overhead (on x-axis) drops to below 3.5$\times$, \sys{} saves energy even for 100\% busy execution; 
when the overhead exceeds 5.2$\times$, \sys{} wastes energy even for 20\% busy execution, the lowest core usage observed on embedded platforms in prior work~\cite{zhai17hotmobile}.
%\note{mention longer execution time will offset benefits} 
%With the XX$\times$ DBT overhead in our baseline design, \sys{} consumes XX\% more energy than the stock Linux!

\paragraph{Qualitative comparison with big.LITTLE}
%We estimate \sys{} saves tangible energy compared to executing device \snr{} on a LITTLE core. 
We estimate \sys{} saves tangible energy compared to a LITTLE core. 
%(same ISA, no offloading needed).
We use parameters based on recent big.LITTLE characterizations~\cite{Hitoshi17PER,Hahnel14HotPower}: 
compared to the big (i.e., CPU on our platform), a LITTLE core has 40 mW idle power~\cite{Peters16CODES} and offers 1.3$\times$ energy efficiency at 70\% clock rate~\cite{Loghin15VLDB}.
%its idle power is 40 mW~\cite{Peters16CODES}.
We favorably assume LITTLE's DRAM utilization is as low as the big, while in reality the utilization should be higher due to LITTLE's smaller LLC. % 512KB
% 23\% redc -- 77?
Even with this favorable assumption for LITTLE and unfavorable, tiny LLC for ARK, LITTLE consumes 77\% 
%(\note{up to XX\%, which should be higher than 77, right?}) 
energy of native execution, more than \sys{} (51\%--66\%), 
%\note{ARK consumes 51\% energy of the native}
%The reason is LITTLE's idle power is only 50\% lower than the CPU;
mainly because LITTLE's idle power is 40$\times$ of Cortex-M3. 
Furthermore, \sys{}'s advantage will be even more pronounced with a proper LLC as discussed earlier. 
%its energy efficiency in busy execution is 5\% lower than \sys{}.
%The key reason is LITTLE almost reduces no energy for idle periods and only sees moderate energy reduction for busy execution. 
%in idle power, e.g., idle waits still consumes 12\% of total energy.
%Therefore its benefit of energy efficiency is much lower than a \pmcore{}.

\begin{comment}
We assume both cores have same DDR bandwidth utilization despite LITTLE core has half of cache size compared to big core.

We use Odroid board with Exynos 5422 Soc as power model.
We use 40 mW and 236mW as LITTLE core's idle wait and busy execution energy consumption, respectively. \note{cite}
%We assume both cores have same memory bandwidth utilization despite LITTLE core has half of cache size compared to big core.
Under this assumption our energy estimation already favors LITTLE core.
The big core is running at 2GHz, and LITTLE core's frequency is 1.4GHz.
Due to micro-architecture difference, big core's performance is around 40\% better compared with LITTLE core under the same frequency. \note{cite}

We estimate the LITTLE core energy using the formula:

$ E_{little} = T_{idle} * P_{idle} + T_{busy} * Freq\_scale * Perf\_scale * P_{busy}$

% Tidle and Tbusy is for execution time on big core
According to our estimation, executing \snr{} on LITTLE core only saves 38\% energy.
This is because a LITTLE core sees limited reduction in idle power, e.g., idle waits still consumes 12\% of total energy.
Therefore its benefit of energy efficiency is much lower than a \pmcore{}.
\end{comment}

% --useful---
%\input{tab-battery-life}

\paragraph{Battery life extension}
Based on \sys{}'s energy reduction in device \snr{}, we project the battery life extension for ephemeral tasks reported in prior work~\cite{drowsy-pm}. 
When the ephemeral tasks are executed at 5-second intervals and the native device \snr{} consumes 90\% system energy in a wakeup cycle, \sys{} extends the battery life by 18\% (4.3 hours per day); 
with 30-second task intervals and a 50\% energy consumption percentage, \sys{} extends the battery life by 7\% (1.6 hours per day). 
This extension is tangible compared to complementary approaches in prior work~\cite{drowsy-pm,central-pm}. 

\subsection{Discussions}
\label{sec:eval:discussion}

%\note{(show the SoC fully conform to the model but the actual implementation needs to be tailored for the hardware)}

\paragraph{Workarounds for OMAP4460}
While OMAP4460 mostly matches our hardware model as summarized in Table~\ref{tab:hwmodel}, 
for minor mismatch we apply the following workarounds. 
\textbfit{Memory mapping}
Our hardware model (\S\ref{sec:bkgnd:hw}) mandates that the \pmcore{} should address the entire kernel memory. 
Yet, Cortex-M3, according to ARM's hardware specification~\cite{m3-trm}, is only able to address memory in certain range (up to \code{0xE0000000}), which 
unfortunately does not encompass the Linux kernel's default address range. 
As a workaround, we configure the Linux kernel source, shifting its address range to be addressable by Cortex-M3.
\textbfit{Interrupt handling}
While our hardware model mandates that both processors should receive all interrupts, OMAP4460 only routes a subset of them (39/102) to Cortex-M3, leaving out IO devices such as certain GPIO pins. 
These IO devices hence are unsupported by the ARK prototype and are not tested in our evaluation.

\paragraph{Recommendation to SoC architects} 
To make SoCs friendly to a \sysmodel{}, architects may consider:
\begin{myenumerateinline}
\item routing all interrupts to CPU and the \pmcore{}, ideally with the identical interrupt line numbers; 
\item making the \pmcore{} capable of addressing the whole memory address space;
\item enlarging the \pmcore{}'s LLC size modestly. 
%The LLC size trades off between the core power and the DRAM power.
We expect a careful increase (e.g., to 64 KB or 128 KB)
%making it on comparable to the DBT code cache, 
will significantly reduce DRAM power at a moderate overhead in the core power.
%If \sys{}'s DRAM utilization is reduced to that of the native execution, \sys{} will reduce 42\% of energy.
%and will significantly reduce the total energy. 
% --- interesting but no space --- %
%In our case, if \pmcore{}'s LlC is large enough to avoid non-necessary threshing(Same memory traffic as CPU), our energy saving will be bumped up from 34\% to 42\%.
%\subsection{What-if analysis}
\end{myenumerateinline}

%\paragraph{Forward compatibility}
%\paragraph{Support new hardware}
%\paragraph{Readiness for new hardware}
%\paragraph{AARCH64: Unsupported}
%\paragraph{Can \sysmodel{} apply to a pair of 64-bit/32-bit ISAs?}
\paragraph{Applicability to a pair of 64-bit/32-bit ISAs}
While today's smart devices often use ARMv7 CPUs, emerging ARM SoCs start to combine 64-bit CPUs (ARMv8) with 32-bit \pmcore{} (ARMv7m), as listed in Table~\ref{tab:hwmodel}. 
On one hand, \sysmodel{}'s idea of exploiting ISA similarity still applies, as exemplified by G2$\rightarrow$H2 in Table~\ref{tab:dbt-aarch64};
on the other hand, its DBT overhead may increase significantly for the following reasons.
Compared to the 32-bit ISA, the 64-bit ISA has richer instruction semantics, more general purpose registers, and a much larger address space.
%\note{below: use ``ARK'' or ``transkernel''?}
As a result, 
\sys{} cannot pass through 64-bit CPU registers but instead have to emulate them in memory; 
\sys{} must translate the guest's 64-bit memory addresses to 32-bit addresses supported by the host (Table~\ref{tab:dbt-aarch64} G1$\rightarrow$H1), 
e.g., by keeping consistent two sets of page tables, for 64-bit and 32-bit virtual address spaces, respectively; 
with large physical memory (>4GB), even this technique will not work because the \pmcore{}'s page tables are incapable of mapping the extra physical memory.

\begin{comment}
Translation overhead of the former two can be mitigated by \sysmodel{} through exploiting their semantic similarity and generating identity mapped instructions, shown in Table \ref{tab:dbt-aarch64} (H2).
Yet, 3) can impose unacceptable overhead due to the lack of effective identity mapping -- 1) \sysmodel{} has to emulate each \textit{x} register (i.e., 64-bit), and 2) provide an emulated page table to mediate \textit{each} 64-bit memory access (H1), assuming the less than 4GB physical memory.
However, when it exceeds 4GB, even software emulation will not work.
\end{comment}

%At present stage \sysmodel{} is bound by incapability to address 64-bit address space and thus does not support AARCH64.
%This will serve as our future work.
% !TeX root = main.tex

% Please add the following required packages to your document preamble:
% \usepackage[table,xcdraw]{xcolor}
% If you use beamer only pass "xcolor=table" option, i.e. \documentclass[xcolor=table]{beamer}
\begin{table}[t]
%\footnotesize
\centering
%\vspace*{-1em}
   	\includegraphics[width=0.48\textwidth]{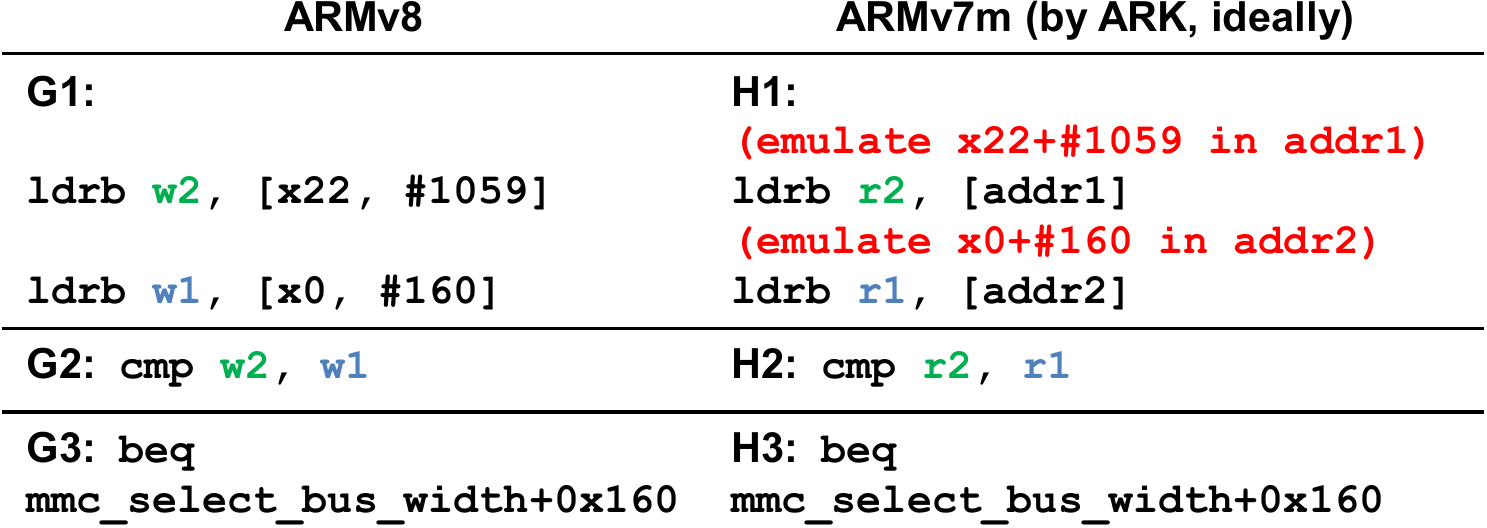}
	\caption{Ideal AARCH64 translation by \sys{} for \code{mmc\_compare\_ext\_csds()} in Linux v4.4.
	While identity mapping still exists (G2$\rightarrow$H2), software emulation can diminish \sys{}'s benefits (G1$\rightarrow$H1).}
	\label{tab:dbt-aarch64}
\vspace*{-1.5em}
\end{table}

%Limitations: v8 vs T32?

\begin{comment} % --- xzl: do it later ---- %

\paragraph{Applicability to other software stack}
\note{be careful -- we said POSIX app is non-goal due to little benefit}
Beyond a monolithic kernel which this work targets, the challenges \sysmodel{} faces can also be found in other software stack -- one has inefficient code not suitable for CPU but is monolithic and complex with changing internal API/ABI, which prevents it from being offloaded quickly to a more energy-efficient core.
Hence in such scenarios, the idea of \sysmodel{} is also applicable and even more inviting (than other approaches): instead of reasoning about an enormous space of changing data types and interfaces, with \sysmodel{} one only has to reason about a few semantically stable APIs and let the DBT handle the rest.
In addition, if such software stack were not system software, \sysmodel{} would be much easier to realize -- it does not have to deal with architectural quirks.

\end{comment}

% !TeX root = main.tex
%\vspace{-5pt}
\section{Related Work}
\label{sec:related}
%\vspace{-10pt}

%\note{add cloud offloading. one sentence in an obvious place}

%\note{from hotmobile. needs to reword!}
%\note{compare to Drowsy}
%\note{cite popcorn2 -- app level support. multi-ISA binary. deep toolchain customization, etc.}

%\sect{related} will discuss power management in firmware.  \note{remove?} \note{merge to below}
 
%\todo{Major parts of related work: 1) PM 2) OS 3) DBT 4) Device Drivers 5) Kernel studies (?) Organize the existing contents and reword}

%x86 and Xeon Phi heterogeneous systems (with distributed OS services?). 

%Shared memory illusion. Highly heterogeneous ISAs.

%Although \sys{} is a novel OS model for programming heterogeneous architecture, it is built upon and inspired by much of the following most related areas of work.

%\note{Liwei: proof this. And cross check with Mothy's comments}

\paragraph{OS for heterogeneous cores}
A multikernel OS runs its kernels on individual processors. 
A number of such OSes are designed anew with a strong distributed system flavor.
They define explicit message interfaces among kernels~\cite{multikernel,fos, m3};
some additionally exploit managed languages/compilers to generate such interfaces~\cite{helios}.  
Unlike them, \sysmodel{} targets spanning an \textit{existing} monolithic kernel and therefore adopts DBT to address the resultant interface challenge.
 
OSes like Popcorn~\cite{popcorn} and K2~\cite{k2} seek to present a single Linux image over heterogeneous processors. 
For sharing kernel state across ISAs, they rely on manual source tweaks or hand-crafted communication. 
They face the interface difficulty as described in \S\ref{sec:bkgnd:sw}.

Prior systems distribute OS functions over CPU and accelerators~\cite{solros,gpufs}. 
The accelerators cannot operate autonomously, which is however required by device \snr{}.
Prior systems offload apps from a smartphone (weak) to cloud servers (strong) for efficiency~\cite{maui,clonecloud}.
Unlike them, \sysmodel{} offloads kernel workloads from a strong processor to a weak \pmcore{} on the same chip. 

\Paragraph{DBT}
%DBT is a classic technique proven useful in virtualization~\cite{qemu}, profiling, 
%%profiling~\cite{valgrind}, 
%and security checks~\cite{kernel-dbt}. 
DBT has been used for system emulation~\cite{qemu} and binary instrumentation~\cite{kernel-dbt,dynamorio,pin,drk}; 
DeVuyst et al.~\cite{devuyst2012asplos} uses DBT to speed up process migration.
Related to \sysmodel{}, prior systems run translated user programs atop an emulated syscall interface~\cite{qemu,hookway1997digital,wang2017enabling}.
Unlike them, \sysmodel{} translates kernel code and emulates a narrow interface \textit{inside} the kernel.
Prior systems use DBT to run binaries in commodity ISAs (e.g., x86) on specialized VLIW cores and hence gain efficiency~\cite{denver,crusoe,hardware-dbt,vliw-dbt}.
None runs on microcontrollers to our knowledge.
\sysmodel{} demonstrates that DBT can gain efficiency even on off-the-shelf cores. 
Existing DBT engines leverage ISA similarities, e.g., between aarch32 and aarch64~\cite{mambo-x64,hypermambo-x64}.
They still fall into the classic DBT paradigm, where the host ISA is brawny and the guest ISA is wimpy (i.e., lower register pressure).
With an inverse DBT paradigm, \sys{} addresses very different challenges. 
%in that the host (where DBT runs) has richer ISA (an aarch64 server, with more general purpose registers) than then guest ISA; by contrast, \sys{} runs a host with a much more constrained ISA and therefore addresses very different challenges. 
%Compared to prior work that runs DBT on powerful, full-fledged hosts, we are the \textbf{first} building DBT for a simple core that even lacks MMU\note{reword}.
%\note{need better discussion:}
%Much work is done on DBT translation rules, with a peephole optimizer~\cite{dbt-peephole}, using machine learning\cite{qemu-ml-rules},  or exploiting the LLVM optimizer~\cite{hqemu}.
Much work is done on optimizing DBT translation rules, using  optimizers~\cite{hqemu,dbt-peephole} or machine learning\cite{qemu-ml-rules}.
Compared to them, \sys{} leverages ISA similarities and hence reuses code optimization already in guest code by guest compilers.

\paragraph{Kernels and drivers}
The \sysmodel{} is inspired by the POSIX emulator~\cite{posix-emulation} however is different as it emulates kernel ABIs.
Prior kernel studies show rapid evolution of the Linux kernel and the interfaces between kernel layers are unstable~\cite{cocinelle1,cocinelle2}. This observation motivates \sysmodel{}. 
Extensive work transplants device drivers to a separate core~\cite{picodriver}, user space~\cite{microdrivers}, or a separate VM~\cite{Levasseur2004}.
%They can be transplanted together with its enclosing kernel to a dedicated core~\cite{picodriver} or VM~\cite{Levasseur2004}.
However, the transplant code cannot operate independent of the kernel, whereas \sysmodel{} must execute autonomously.

%http://lib.tkk.fi/Diss/2012/isbn9789526049175/isbn9789526049175.pdf, page 198
%https://archive.fosdem.org/2013/schedule/event/operating_systems_anykernel/
%https://archive.fosdem.org/2014/schedule/event/01_uk_rump_kernels/

Encapsulating the NetBSD kernel subsystems (e.g., drivers) behind stable interfaces respected by developers, 
rump kernel~\cite{rumpkernel} seeks to enable their reuse in foreign environments, e.g., hypervisors. 
The \sysmodel{} targets a different goal: spanning a live Linux kernel instance over heterogeneous processors. 
Applying Rump kernel's approach to Linux is difficult, as Linux intentionally rejects interface stability for drivers ~\cite{stable-api-nonsense}.

\noindent
\textbf{Suspend/resume}'s
inefficiency raises attention for cloud servers~\cite{powernap,pm-critical-path} and mobile~\cite{drowsy-pm}.
%The significance of efficient suspend/resume has been recognized for 
%cloud servers~\cite{powernap} and 
%mobile devices~\cite{drowsy-pm}. 
%Prior work has already shown some evidences that IO devices are one key bottleneck
Drowsy~\cite{drowsy-pm} mitigates inefficiency by reducing the devices involved in \snr{} through user/kernel co-design; Xi \textit{et al.} propose to reorder devices to resume~\cite{pm-critical-path}.
While acknowledging the value of such kernel optimizations, we believe ARK is a key complement that works on unmodified binaries. 
ARK can co-exist with the mentioned optimizations in the same kernel. 
%Our approach is complementary and requires no changes to the userspace. 
PowerNap~\cite{powernap} takes a hardware approach to speed up \snr{} for servers.
It does not treat kernel execution for operating diverse IO on embedded platforms.
Kernels may put idle devices to low power at runtime~\cite{central-pm},  complementary to \snr{} that ensures all devices are off.

%On one hand, prior work seeks to improve efficiency by reducing the energy of each suspend/resume~\cite{drowsy-pm,pm-critical-path};
%complementary to it, our approach aims to reduce the energy cost by exploiting the ISA similarity exemplified by increasingly emerging platforms. 
%On the other hand, prior work\cite{central-pm} advocates new power management scheme, such as automatic \textit{runtime} IO power management or hardware-assisted suspend/resume~\cite{powernap}; orthogonal to it, this work faces non-latency-sensitive tasks and address diverse IO on embedded platforms.

%: 1) we serve background tasks (longer delay is acceptable); 2) we focus on the OS execution in the transitions; 3) we address the diverse IO on embedded platforms.
% !TeX root = main.tex

%\vspace{-5pt}
\section{Conclusions}
%\vspace{-5pt}
We present \sysmodel{}, a new executor model for a \pmcore{} to execute a commodity kernel's phases, notably device \snr{}. 
The \sysmodel{} executes the kernel binary through cross-ISA DBT. 
It translates stateful code while emulating stateless services; 
it picks a stable ABI for emulation; 
it specializes for hot paths; 
it exploits ISA similarities for DBT. 
%Through a concrete implementation, 
We experimentally demonstrate that the approach is feasible and beneficial. 
The \sysmodel{} represents a new OS design point for harnessing heterogeneous SoCs.

% --- 11 page --- %
% !TeX root = main.tex

\section*{Acknowledgments}
The authors were supported in part by NSF Award \#1619075, \#1718702, and a Google Faculty Award. 
The authors thank the paper shepherd, Prof. Timothy Roscoe, and the anonymous reviewers for their insightful feedback. 
The authors thank Prof. Lin Zhong for providing a JTAG debugger. 
%for OMAP4460.
%from Rice University for his JTAG debugger, which makes the nightmare of debugging gentler.
The authors are grateful to numerous video game emulators that inspired this project. 
%Transkernel was inspired by numerous video game emulators. % (VirtuaNES and ZSNES). 
%that built atop binary translation. 

%\footnotesize 
%\bibliographystyle{acm}
%\bibliography{../common/bibliography}}
\bibliographystyle{bib/abbrv-minimal}
%\bibliography{bib/abr-short,bib/xzl,bib/PM,bib/iovirt,bib/misc,bib/book,bib/dbt,bib/iot}
\bibliography{bib/abr-long,bib/export}
%\theendnotes

\end{document}